\documentclass[manuscript]{aastex63}

\usepackage{natbib}

\newcommand{\cpld}{cyclopropenylidene}
\newcommand{\pdld}{propadienylidene}
\newcommand{\ppld}{propynylidene}

\newcommand{\ccchh}{C$_3$H$_2$}
\newcommand{\etcyn}{C$_2$H$_5$CN}
\newcommand{\vycyn}{C$_2$H$_3$CN}
\newcommand{\methcyn}{CH$_3$CN}
\newcommand{\pyrid}{C$_5$H$_5$N}
\newcommand{\pyrim}{C$_4$H$_4$N$_2$}
\newcommand{\benzene}{C$_6$H$_6$}
\newcommand{\cyanogen}{C$_2$N$_2$}
\newcommand{\methacet}{C$_3$H$_4$}
\newcommand{\diacet}{C$_4$H$_2$}
\newcommand{\hcccn}{HC$_3$N}
\newcommand{\propane}{C$_3$H$_8$}
\newcommand{\coo}{CO$_2$}
\newcommand{\allene}{CH$_2$CCH$_2$}
\newcommand{\propene}{C$_3$H$_6$}
\newcommand{\water}{H$_2$O}
\newcommand{\hydrogen}{H$_2$}
\newcommand{\nitrogen}{N$_2$}
\newcommand{\methane}{CH$_4$}
\newcommand{\ethylene}{C$_2$H$_4$}
\newcommand{\ethane}{C$_2$H$_6$}
\newcommand{\acet}{C$_2$H$_2$}

\newcommand{\dg}{$^{\circ}$}

\received{22 April 2020}
\revised{14 August 2020}
\accepted{7 September 2020}

\submitjournal{The Astronomical Journal}

\shorttitle{Detection of Cyclopropenylidene  on Titan}
\shortauthors{Nixon et al.}

\begin{document}

\title{Detection of Cyclopropenylidene  on Titan with ALMA}

\correspondingauthor{Conor A. Nixon}
\email{conor.a.nixon@nasa.gov}

\author[0000-0001-9540-9121]{Conor A. Nixon}
\affiliation{Solar System Exploration Division, NASA Goddard Space Flight Center, Greenbelt, MD 20771, USA.}

\author[0000-0002-8178-1042]{Alexander E. Thelen}
\altaffiliation{NASA Post-Doctoral Program Fellow}
\affiliation{Universities Space Research Association, Columbia, MD 21046, USA}
\affiliation{Solar System Exploration Division, NASA Goddard Space Flight Center, Greenbelt, MD 20771, USA.}

\author[0000-0001-8233-2436]{Martin A. Cordiner}
\affiliation{Catholic University of America, Washington, DC 20064, USA}
\affiliation{Solar System Exploration Division, NASA Goddard Space Flight Center, Greenbelt, MD 20771, USA.}

\author[0000-0002-2570-3154]{Zbigniew Kisiel}
\affiliation{Institute of Physics, Polish Academy of Sciences, Al. Lotnik{\'o}w 32/46, 02-668 Warszawa, Poland.}

\author[0000-0001-6752-5109]{Steven B. Charnley}
\affiliation{Solar System Exploration Division, NASA Goddard Space Flight Center, Greenbelt, MD 20771, USA.}

\author[0000-0003-3799-9033]{Edward M. Molter}
\affiliation{Department of Astronomy, University of California, Berkeley, CA 94720, USA.}

\author[0000-0003-3849-5064]{Joseph Serigano}
\affiliation{Department of Earth and Planetary Sciences, Johns Hopkins University, Baltimore, MD 21218, USA.}

\author[0000-0002-6772-384X]{Patrick G. J. Irwin}
\affiliation{Atmospheric, Oceanic, and Planetary Physics, Clarendon Laboratory, University of Oxford, Parks Road, Oxford, OX1 3PU, UK.}

\author[0000-0003-3108-5775]{Nicholas A. Teanby}
\affiliation{School of Earth Sciences, University of Bristol, Wills Memorial Building, Queens Road, Bristol, BS8 1RJ, UK.}

\author[0000-0002-4336-0730]{Yi-Jehng Kuan}
\affiliation{Department of Earth Sciences, National Taiwan Normal University, Taipei 116, Taiwan, ROC.}
\affiliation{Institute of Astronomy and Astrophysics, Academia Sinica, Taipei 106, Taiwan, ROC.}




\begin{abstract}


We report the first detection on Titan of the small cyclic molecule \cpld\ (c-\ccchh ) from high sensitivity spectroscopic observations made with the Atacama Large Millimeter/sub-millimeter Array (ALMA). Multiple lines of \cpld\ were detected in two separate datasets: $\sim$251 GHz in 2016 (Band 6) and $\sim$352 GHz in 2017 (Band 7). Modeling of these emissions indicates abundances of 0.50 $\pm$ 0.14 ppb (2016) and 0.28 $\pm$ 0.08 (2017) for a 350-km step model, which may either signify a decrease in abundance, or a mean value of 0.33 $\pm$ 0.07 ppb. Inferred column abundances are 3--5  $\times 10^{12}$~cm$^{-2}$ in 2016 and 1--2  $\times 10^{12}$~cm$^{-2}$ in 2017, similar to photochemical model predictions.
Previously the C$_3$H$_3^+$ ion has been measured in Titan's ionosphere by Cassini's Ion and Neutral Mass Spectrometer (INMS), but the neutral (unprotonated) species has not been detected until now, and aromatic versus aliphatic structure could not be determined by the INMS. Our work therefore represents the first unambiguous detection of \cpld ,  the second known cyclic molecule in Titan's atmosphere along with benzene (\benzene ) and the first time this molecule has been detected in a planetary atmosphere. We also searched for the N-heterocycle molecules pyridine and pyrimidine finding non-detections in both cases, and determining 2-$\sigma$ upper limits of 1.15 ppb (c-\pyrid ) and 0.85 ppb (c-\pyrim ) for uniform abundances above 300~km.
These new results on cyclic molecules provide fresh constraints on photochemical pathways in Titan's atmosphere, and will require new modeling and experimental work to fully understand the implications for complex molecule formation.

\end{abstract}

\keywords{TBD}


\section{Introduction} 
\label{sect:intro}

Saturn's moon Titan exhibits the most complex chemistry of any known planetary atmosphere other than the Earth. The reducing chemical environment, composed primarily of methane and nitrogen gases \citep{niemann10}, produces a rich array of organic molecules when activated by solar UV photons or Saturn magnetospheric electrons \citep{vuitton19}. Many of these daughter species are hydrocarbons (C$_x$H$_y$) or nitriles (C$_x$H$_y$(CN)$_z$), however several oxygen compounds have also been detected (CO, \coo , \water ), apparently due to an influx of external OH and O$^+$ from Enceladus \citep{horst08}, and several other light gases including \hydrogen\ (from methane destruction), and the noble gases Ar and Ne.

Prior to the Cassini mission, most of our knowledge about Titan's atmospheric composition had come from remote sensing spectroscopy. While \methane\ and \nitrogen\ were detected at short wavelengths \citep{kuiper44, broadfoot81}, most other gases were first seen in the infrared. These include the  detections of \ethane , \acet , \ethylene\  and CO using ground-based telescopes \citep{gillett73, gillett75, lutz83}; the Voyager 1 IRIS (Infrared Interferometer-Spectrometer) detections of \hydrogen , \methacet , \propane , \diacet ,  HCN, \hcccn , \cyanogen\ and \coo\ \citep{hanel81, maguire81, kunde81, samuelson81, samuelson83}; as well as later detections with ISO (Infrared Space Observatory) of \water\ and \benzene\ \citep{coustenis98, coustenis03}. A notable exception was the detection of \methcyn\ by \citet{bezard92} at sub-millimeter wavelengths using the IRAM 30~m telescope at Pico Veleta.

This paradigm changed substantially with the Cassini-Huygens mission, which carried mass spectrometers on both the orbiter and the probe \citep{niemann02, young04, waite04}, able to sample the composition of Titan's atmosphere {\em in situ} for the first time. Modeling of these mass spectra revealed a plethora of ion and neutral species \citep{[e.g.][]waite05, hartle06, vuitton07, vuitton09, cui09, bell10a, bell10b, westlake11}, although in many cases exact molecular identification remained elusive, due to the inability of mass spectra alone to elucidate molecular structure. One new positive identification was made in the infrared using Cassini's CIRS instrument \citep[Composite Infrared Spectrometer,][]{flasar04a}, of propene \citep[\propene ,][]{nixon13a}. Shortly after the end of the Cassini mission, a further infrared detection was made using TEXES (the Texas Echelon-cross-Echelle Spectrograph) \citep{lacy02} at NASA's Infrared Telescope Facility (IRTF): namely propadiene \citep[\allene ,][]{lombardo19c}, an isomer of propyne (\methacet ).

The newest tool for probing Titan's atmospheric composition has been ALMA \citep[the Atacama Large Millimeter/submillimeter Array:][]{baars02, lellouch07}, a powerful interferometer array that started science observations in 2011. At millimeter (mm) and submillimeter (sub-mm) wavelengths rotational transitions of molecules are accessible, which have proved vital for probing the chemistry of astrophysical objects such as dense molecular clouds. Using early data from ALMA two further nitrile (cyanide) species were soon conclusively identified in Titan's atmosphere: propionitrile \citep[ethyl cyanide, \etcyn ,][]{cordiner15} and acrylonitrile \citep[vinyl cyanide, \vycyn ,][]{palmer17}, as well as many isotopologues of previously detected species including CO, HCN, \hcccn , \methcyn\ and \methane\ \citep[][]{serigano16, molter16, palmer17, cordiner18, thelen19a, iino20}.

Besides making new chemical detections, observations of Titan from Cassini and ALMA have mapped the spatial and temporal evolution of the gas distributions, revealing complex structure such as polar jets, and seasonal changes of unexpected rapidity: see \cite{bezard14, horst17} for detailed reviews. In parallel with observations, photochemical modeling of Titan's atmosphere has also progressed rapidly to explain the observed gas abundance distributions, and to make predictions for target species likely to be detectable. See for example recent work by \citet{krasnopolsky09, krasnopolsky10, krasnopolsky12, hebrard13, krasnopolsky14, dobrijevic14, loison15, willacy16, vuitton19}. 

In 2016 and 2017 we conducted high sensitivity observations with ALMA, with the goal of searching for new molecules in Titan's atmosphere, including the N-heterocyclic molecules pyridine (c-\pyrid ) and pyrimidine (c-\pyrim ). N-heterocycles have a strong importance to astrobiology since these form the backbone rings for the nucleobases of DNA and RNA. Neither of these molecules were detected, and upper limits on their abundances determined instead. However, we did make a first detection on Titan of \cpld , a small cyclic hydrocarbon molecule that has previously been detected in astrophysical sources but not in a planetary atmosphere.

This paper is organized as follows. In Section \ref{sect:obs} we describe the observations and data reduction, and in Section \ref{sect:modeling} the data modeling process. In Section \ref{sect:results} we show the results, followed by a discussion in Section \ref{sect:disc} and conclusions in Section \ref{sect:conc}. 

\section{Observations} 
\label{sect:obs}

Observations of Titan were completed during March 2--4 2016 in Band 6 (ALMA Project Code 2015.1.00423.S) and on May 8th and 16th 2017 in Band 7 (ALMA Project Code 2016.A.00014.S), see Table~\ref{tab:obs}. In addition, part of a third dataset was used to obtain a CO J=2${\rightarrow}$1 observation of Titan in 2016 for retrieval of the disk-averaged temperature profile. In this independent dataset (ALMA 2015.1.00512.S, observed April 1st 2016) Titan was observed as a flux calibration target for an astrophysical investigation. Details of spectral windows (Spw) analyzed in this paper are given in Table~\ref{tab:windows}. 

For dataset 2015.1.00423.S the data were provided in calibrated form (bandpass, phase and flux calibrated), and subsequently post-processed using the Common Astronomy Software Applications (CASA) package Version 4.7.2-REL (r39762, March 8th 2017) to provide rest-velocity correction ({\tt cvel}) and ephemeris update ({\tt fixplanets}). Lastly the data were concatenated and then deconvolved (`cleaned') in CASA using the H{\"o}gbom algorithm, with a cell size of 0.1$\arcsec$ and a threshold of 10 mJy, and a final restoring beam size of 0.87$\arcsec \times$0.72$\arcsec$

For the 2016 CO dataset (ALMA 2015.1.00512.S) the data were reduced in CASA Version 5.6.1-8 using the ALMA pipeline script prepared by the Joint ALMA Observatory staff, with the exception of the removal of the {\tt {hifa\_fluxcalflag}} task so that Titan's atmospheric CO J=2${\rightarrow}$1 emission line at 230538 GHz was not flagged out. Data were deconvolved with the CASA {\tt clean} task, using the H{\"o}gbom algorithm with an image size = 128$\times$128 pixels, where pixels were set to 0.2$\arcsec \times$0.2$\arcsec$. The resulting synthesized beam had FWHM (Full Width to Half Maximum) = 0.92$\arcsec \times$0.81$\arcsec$, comparable to Titan's angular size at the time of observing.

The data reduction of the 2017 data (2016.A.00014.S) has already been described in \citet{cordiner19}. In addition, the bandpass solution interval was increased to 10 channels (2.44~MHz) to further improve the S/N and aid in the detection of weak spectral lines (Yamaki et al., 2012).

\begin{table}[ht]
\caption{Details of ALMA Observations Of Titan}
\begin{centering}
\footnotesize
\begin{tabular}{cccccccccc}
 & & & & & & & & & \\
\hline
Date & Start  & End  & $t$~*  
& $\Delta v$ $^\dagger$  & $\Delta f$ $^\ddagger$ & Beam  & Position & Angular & Sub-Earth \\
 & (UT) & (UT) & (mins) & (km s$^{-1}$) & (MHz)  & Size & Angle & Diameter ($\arcsec$) & Latitude \\
 \hline
 \multicolumn{10}{c}{\em Project Code 2015.1.00423.S} \\
 02-Mar-2016  & 09:44 & 10:57 & 43 & -28.37 & 24.13  & 0.87$\arcsec \times$0.75$\arcsec$ & 89.145\dg & 0.71 & 26.28 \\
 02-Mar-2016  & 11:03 & 12:14 & 43 & -28.34 & 24.01  & 0.91$\arcsec \times$0.73$\arcsec$ & 95.155\dg & 0.71 & 26.28 \\
 04-Mar-2016  & 09:43 & 10:55 & 43 & -32.18 & 27.37  & 0.87$\arcsec \times$0.64$\arcsec$ & 79.815\dg & 0.71 & 26.28 \\
\multicolumn{10}{c}{\em Project Code 2015.1.00512.S} \\
01-Apr-2016 & 08:22 & 08:24 & 2 & -22.32 & 17.12 &  0.92$\arcsec \times$0.81$\arcsec$ &  -83.247\dg & 0.74 & 26.24 \\
\multicolumn{10}{c}{\em Project Code 2016.A.00014.S} \\ 
08-May-2017 & 08:49 & 09:07 & 18 & -19.28 & 22.24 & 0.18$\arcsec \times$0.15$\arcsec$ & -80.950\dg & 0.77 & 26.41 \\
16-May-2017 & 05:34 & 07:12 & 98 & -13.61 & 15.97 & 0.28$\arcsec \times$0.19$\arcsec$ & -73.549\dg & 0.77 & 26.48 \\
\hline
 & & & & & & & & &  \\
\end{tabular}
\end{centering}
\normalsize
\newline
\hspace*{3cm} * Time spent on source. \newline 
\hspace*{3cm} $^{\dagger}$ Topocentric velocity (negative = approaching). \newline  
\hspace*{3cm} $^{\ddagger}$ Frequency doppler shift (positive = blue-shifted). 
\label{tab:obs}
\end{table}

\begin{table}[ht]
\caption{Details of observation spectral windows}
\begin{tabular}{ccccccc}
\hline
Spw & Freq. Range (MHz) & $\Delta f ^{\dagger}$ (MHz) & $n_{\rm channels}$ & Molecule & $f_0$ (MHz) & Transition \\
\hline
\multicolumn{7}{c}{\em Project Code 2015.1.00423.S} \\
0 & 249570--250050 & 0.244 & 1920 & c-\pyrim\ & 249820 & $J"$=39, $^bR$-band \\
1 & 251260--251740 & 0.244 & 1920 & c-\pyrid\ & 251510 & $J"$=41, $^aR$-band \\
2 & 261900--262380 & 0.244 & 1920 & c-\pyrim\ & 262150 & $J"$=41, $^bR$-band \\
3 & 263090--263570 & 0.244 & 1920 & c-\pyrid\ & 263340 & $J"$=43, $^aR$-band \\
\multicolumn{7}{c}{\em Project Code 2015.1.00512.S} \\
4 & 230322 - 230791& 0.244 & 1920 & CO & 230538 & $J=2\rightarrow$1 \\
\multicolumn{7}{c}{\em Project Code 2016.A.00014.S} \\
5 & 344212-346085 & 0.977 & 1920 & CO & 345796 & $J=3\rightarrow$2 \\
6 & 351281-352219 & 0.244 & 3840 & {\vycyn} & --- & multiple \\
\hline
\end{tabular}
\newline
\newline
\hspace*{2cm} $^{\dagger}$Channel spacing: spectral resolution is twice the channel spacing.
\label{tab:windows}
\end{table}

Disk-averaged spectra from all observations were extracted from an integrated region defined by a circular pixel mask set to contain 90\% of Titan's continuum flux, as in \citet{lai17}.

\section{Modeling} 
\label{sect:modeling}

Modeling was accomplished using the NEMESIS program \citep[][]{irwin08}, which has previously been successfully applied to model ALMA spectra of Titan \citep[e.g.][]{cordiner15,molter16,serigano16,palmer17,lai17,teanby18, thelen18, thelen19a, thelen19b}. The NEMESIS fitting algorithm uses a Bayesian optimal estimation technique as described by \citet{rodgers00}, which seeks to minimize a `cost function' similar to a $\chi^2$ figure of merit,  which penalizes the solution according to the square deviation of both the solution vector from the original {\em a priori} state, and also the model spectrum from the data.  Marquart-Levenberg minimization is used to descend a downhill trajectory of the cost function until satisfactory convergence is reached (solution changing by $<$0.1\%). ALMA spectra were rescaled to radiance units before being input to NEMESIS, and then modeled using a weighted average of spectra calculated at 35 emission angles from disk center to 1200 km altitude (3775 km radius), as described in \citet{teanby13}, Appendix~A. 

Spectral line data for most molecules were taken from the CDMS catalog \citep[][\tt https://cdms.astro.uni-koeln.de]{muller01,muller05}, which is a compilation of transition information from the published literature. These include: HCN \citep{ahrens02, fuchs04, cazzoli05, maiwald00}, CO \citep{winnewisser97, goorvitch94}, \methcyn\ \citep{kukolich73, kukolich82, boucher77, cazzoli06} , \vycyn\ \citep{muller08}, c-\pyrim\ \citep{kisiel99} and c-\ccchh\ \citep{bogey86, vrtilek87, lovas92}, including their isotopes. The rotational spectrum of c-\pyrid\ was calculated by refitting the primary spectroscopic data from \citet{heineking86, wlodarczak88}. For \etcyn , we used a new, more complete spectral line list that included not only rotations in the ground vibrational state, but also in the first three vibrational states as described in \citet{kisiel20}. 

Collision-induced opacity for relevant molecular pairs was computed using published formalisms and publicly-available FORTRAN codes as follows: 
\nitrogen --\nitrogen\ \citep{borysow86c}; 
\nitrogen --\methane\ \citep{borysow93}; 
\methane --\methane\  \citep{borysow87}; 
\nitrogen --\hydrogen\ \citep{borysow86a};  
\methane --\hydrogen\  \citep{borysow86b}; 
\hydrogen --\hydrogen\ \citep{borysow91}.

\subsection{Temperature Retrievals} 

\begin{figure}
\includegraphics[width=18cm]{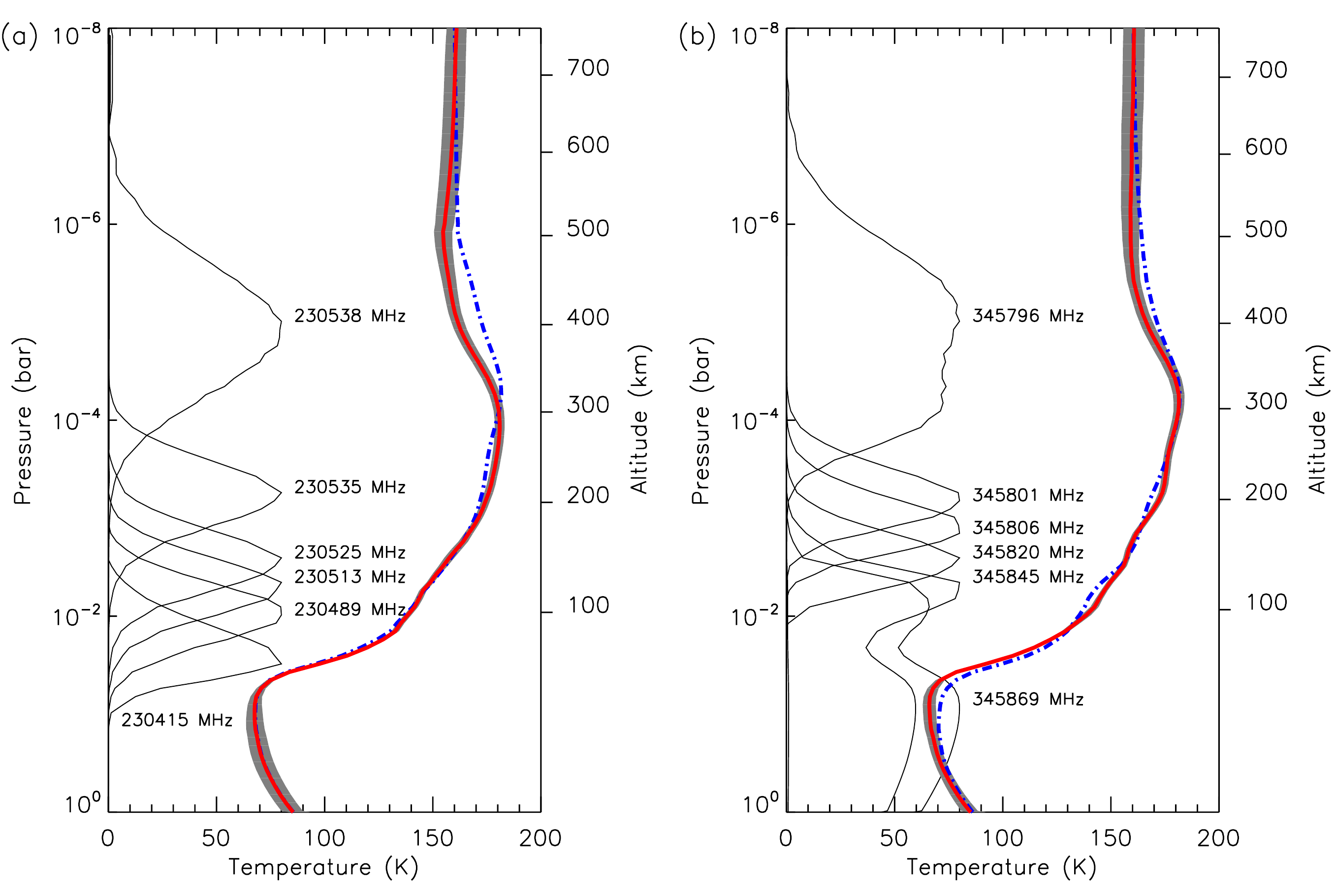}
\caption{ 
Temperature retrievals for (a) CO 2$\rightarrow$1 in 2016 and (b) CO 3$\rightarrow$2 in 2017. 
Blue dot-dash lines: {\em a priori} temperature profiles. Red solid line: retrieved profiles, with shaded (grey) retrieval error  indicated. Normalized contribution functions at different frequencies are also shown (thin solid black lines).
\label{fig:temps}}
\end{figure}

Firstly, the spectral lines of CO (Spw 4 \& 5) were fitted using a model that allowed continuous variation of the temperature profile between 100--500~km, while CO was fixed at a constant mixing fraction of 49.6~ppm as determined by \citet{serigano16}.  The {\em a priori} temperature profile was constructed by interpolating measurements from the Huygens Atmospheric Structure Instrument (HASI) and Cassini radio science observations to Titan's sub-observer latitude ($\sim$26\dg ) below 100~km \citep{fulchignoni05, schinder12}, and disk-averaged retrieval results from 2015 ALMA observations of Titan \citep{thelen18} were used at altitudes $>$100 km. Temperature {\em a priori} errors were set to 5~K in all atmospheric layers (0--1200 km), which allowed NEMESIS to obtain a fit to the data while limiting artificial vertical structure (ill-conditioning) in the retrieved temperature profile. Different frequency offsets from the line center sounded to different atmospheric depths (altitudes, or pressure levels), as shown by the contribution functions in Fig.~\ref{fig:temps}. The {\em a priori} and final retrieved temperature profiles are also indicated.

\subsection{Spectral windows 1 and 6: Discovery of Cyclopropenylidene}
\label{sect:win16}

Next Spw 1\footnote{Window 0 was not included in this study. This region contains isotopic lines of CH$_3$C$^{15}$N, the analysis of which will be reported elsewhere. See also recent work on this isotopologue by \citet{iino20}} was modeled to fit visible lines of known molecules: \etcyn\ and \vycyn . The temperature profile was fixed at the earlier retrieved profile from Spw 4 for 2016. Various gas profile types were investigated for the nitriles, adjusting the profiles to achieve best fits. 

\begin{figure}
\includegraphics[width=17cm]{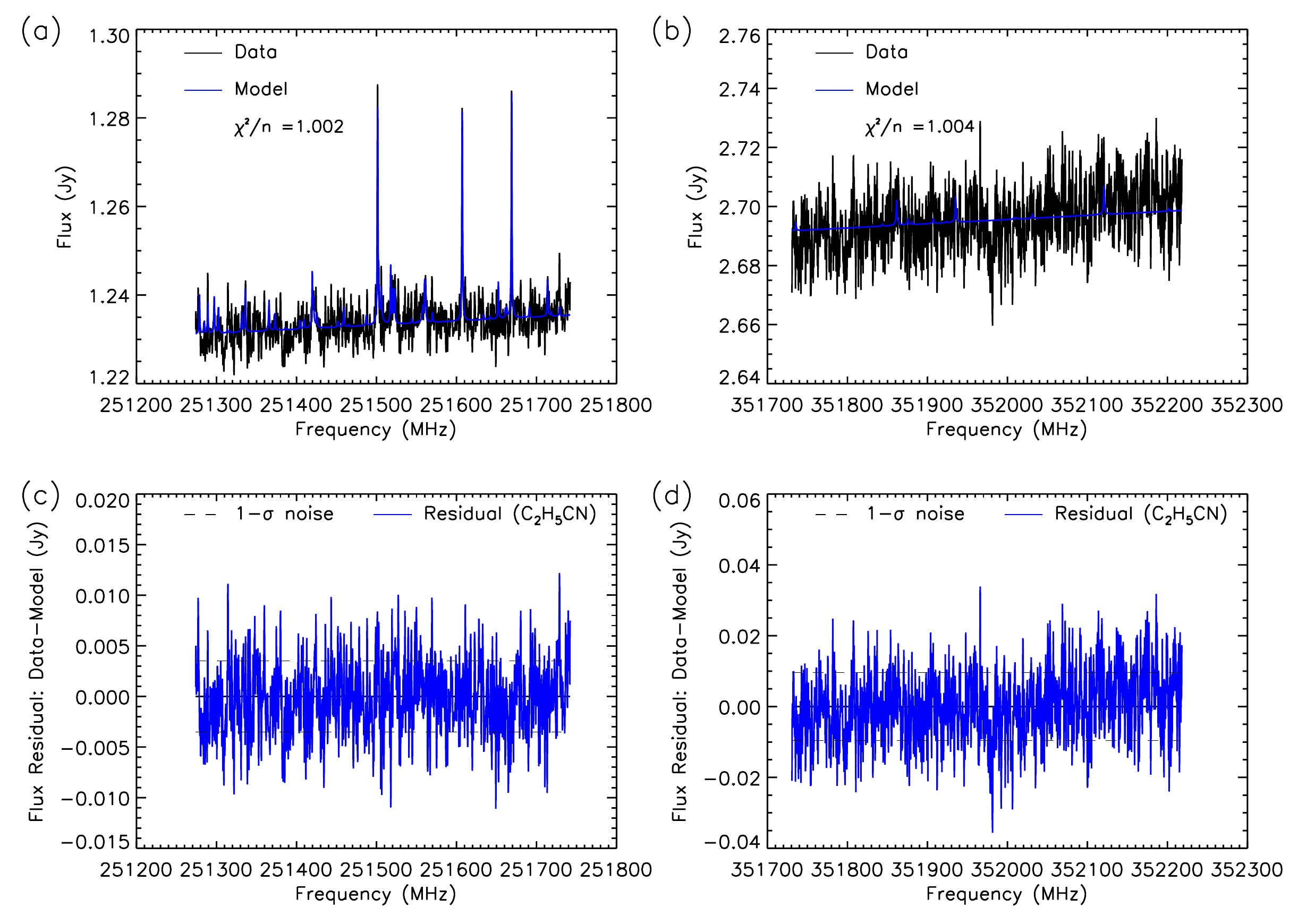}
\caption{ 
Best fit models for Spw 1 (a) and Spw 6 (b) using previously known gases only. Gas profiles were step function models for \etcyn\ (250~km) and \vycyn\ (300~km). Residuals after fitting are shown in panels (c) and (d). Frequencies have been corrected to rest velocity frame.
}
 \label{fig:win16resid}
\end{figure}

We first tried using minimalist `step' functions (uniform volume mixing ratio above a fixed pressure level, and zero below) for the vertical distribution of each gas. From previous experience \citep[e.g.][]{cordiner15,lai17} we found that these worked well for trace (low abundance) nitriles in the ALMA spectrum where there is little information that can be obtained about the vertical profile. For \vycyn\ we adopted a step altitude at 300 km. For \etcyn , there was sufficient sensitivity to the altitude of the step to affect the quality of the fit, which was determined from Spw 1 and thereafter fixed at 250 km (see Appendix~\ref{sect:c2h5cn}). Initial fitting is shown in Fig.~\ref{fig:win16resid}.

Having fitted the features of the known nitrile gases as well as possible, we proceeded to try adding additional gases to the model in an attempt to detect any weak lines due to a new species, including {\em i}-butanenitrile and {\em n}-butanenitrile (C$_3$H$_7$CN), propynenitrile (cyanodiacetylene, HC$_5$N) and others, using lines from the JPL catalog \citep[][\tt https://spec.jpl.nasa.gov]{pickett98}. In both Spw 1 and 6, we found a significant improvement to the model fit after introducing the gas c-\ccchh\ (\cpld ) using spectroscopic lines from CDMS originally determined by \citet{bogey86, vrtilek87} with a trial step function model with a step at 300~km, or higher. 

\begin{figure}
\includegraphics[width=18cm]{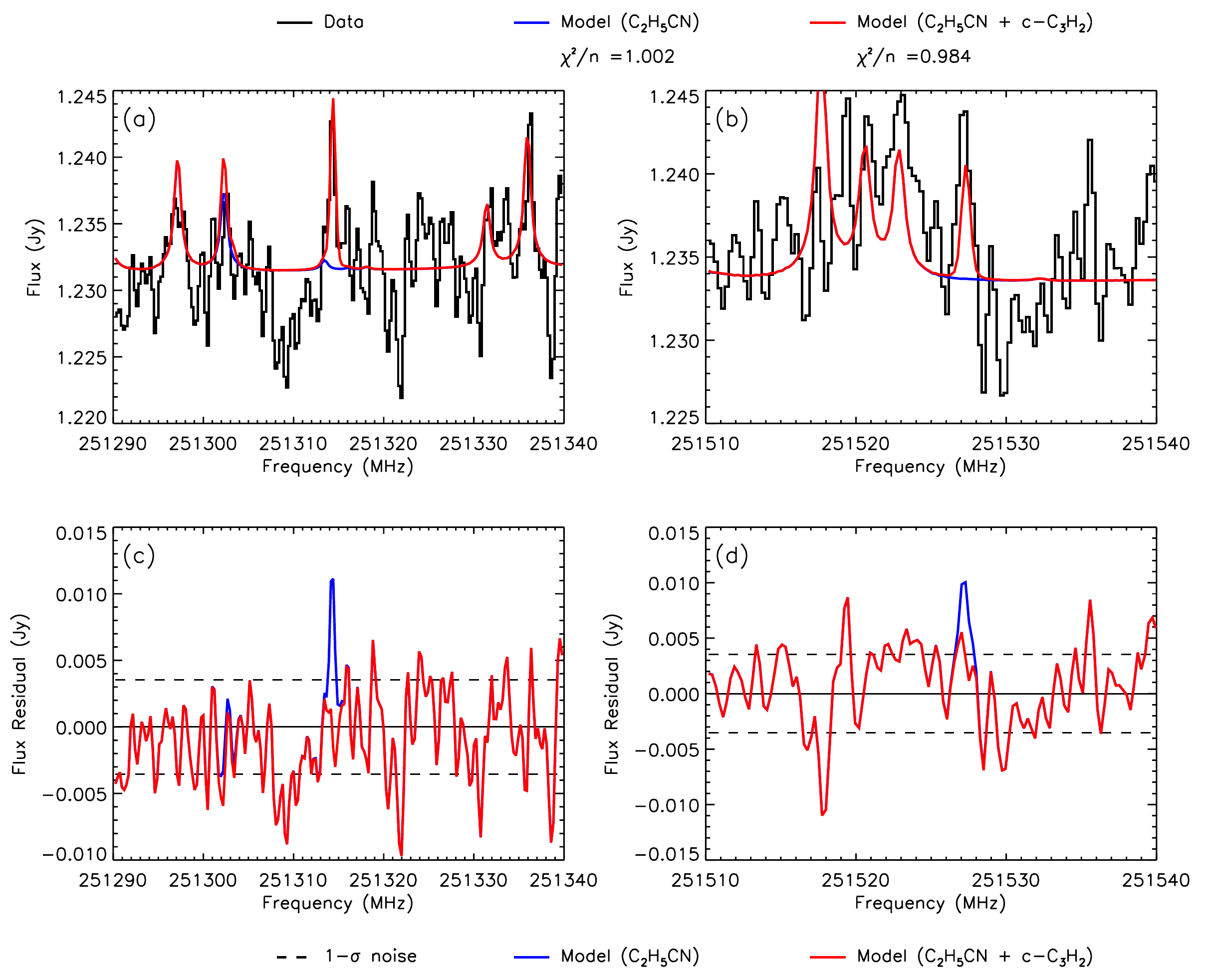}
\caption{ 
Modeling of Spw 1 showing expanded scale of regions where detected c-\ccchh\ lines are present: blend of $7_{0, 7} \rightarrow 6_{1, 6}$ and $7_{ 1, 7} \rightarrow  6_{0, 6}$ transitions at 251314.3 MHz, and $6_{2, 5} \rightarrow 5_{1, 4}$ single transition at 251527.3 MHz. Blue: \etcyn\ model only. Red: model with \etcyn\ and c-\ccchh . Frequencies have been corrected to rest velocity frame.
 \label{fig:c3h2-2016-zoom}}
\end{figure}

In Spw 1 two significant lines were detected at 251314.3 MHz (blend of $7_{0, 7} \rightarrow 6_{1, 6}$ and $7_{ 1, 7} \rightarrow  6_{0, 6}$ transitions) and 251527.3 MHz ($6_{2, 5} \rightarrow 5_{1, 4}$), as shown in Fig.~\ref{fig:c3h2-2016-zoom}. We note that these two emissions are the strongest expected spectral features of c-\ccchh\ in Spw 1, and show close to the expected proportions of relative intensities. 


\begin{figure}
\includegraphics[width=18cm]{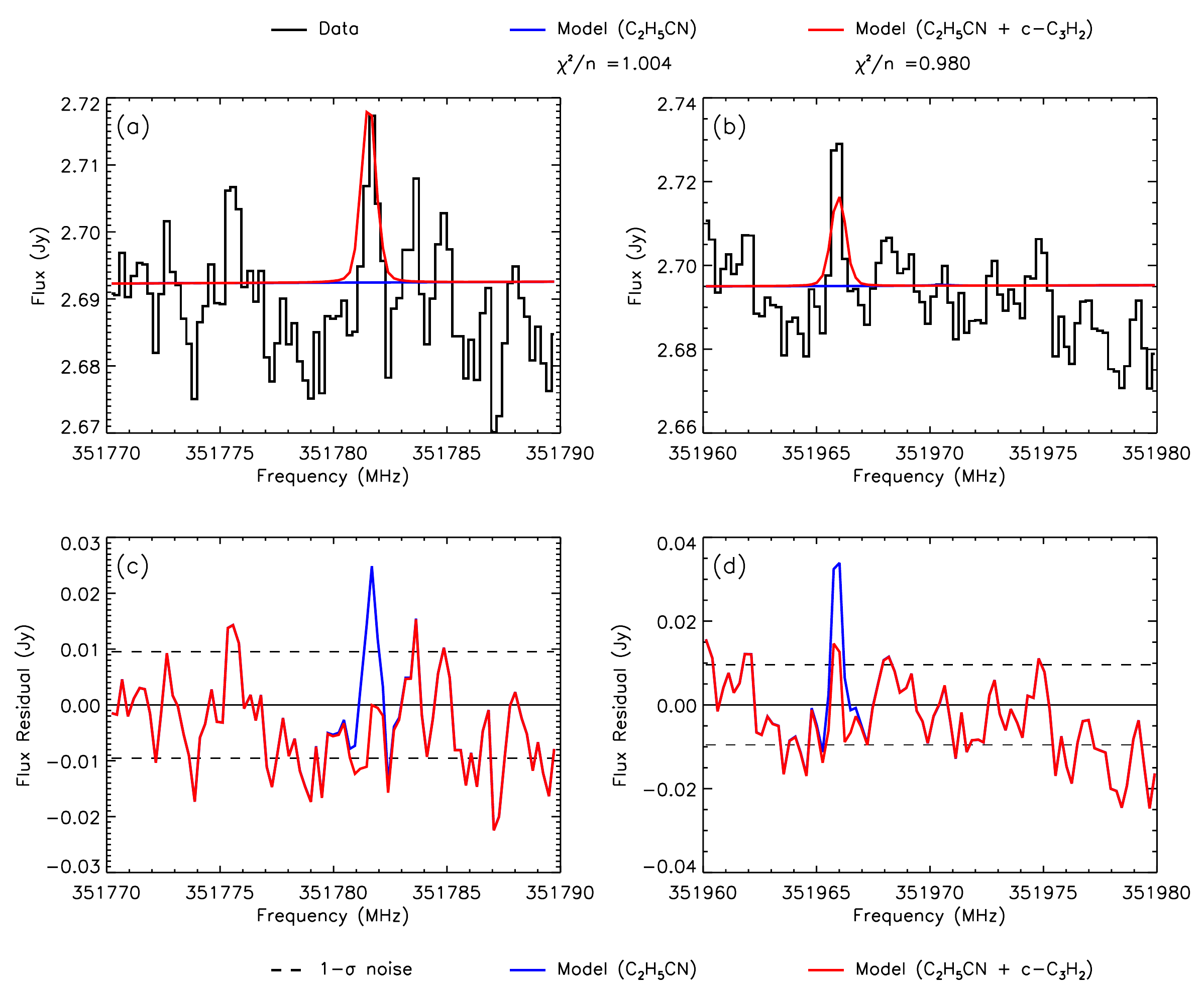}
\caption{ 
Modeling of Spw 6 showing expanded scale of regions where detected c-\ccchh\ lines are present: blend of $10_{1, 10} \rightarrow 9_{0, 9}$ and $10_{ 0, 10} \rightarrow  9_{1, 9}$ doublet at 351781.6 MHz, and blend of $9_{1, 8} \rightarrow 8_{2, 7}$ and $9_{ 2, 8} \rightarrow  8_{1, 7}$ doublet at 351965.9 MHz. Blue: \etcyn\ model only. Red: model with \etcyn\ and c-\ccchh . Frequencies have been corrected to rest velocity frame.
 \label{fig:c3h2-2017-zoom}}
\end{figure}

Similarly, in Spw 6, despite the noise level being higher in ALMA Band 7 than in Band 6 (Spw 1), we made two further detections of lines of c-\ccchh : 
351781.6 MHz (blend of the $10_{1, 10} \rightarrow 9_{0, 9}$ and $10_{ 0, 10} \rightarrow  9_{1, 9}$ doublet),
351965.9 MHz (blend of $9_{1, 8} \rightarrow 8_{2, 7}$ and $9_{ 2, 8} \rightarrow  8_{1, 7}$ doublet),
see Fig.~\ref{fig:c3h2-2017-zoom}.

To further test the detection of c-\ccchh , we calculated a $\Delta \chi^2$ curve for different amounts of the gas in a forward model, using a step function at 350 km. In this case, $\chi^2 = \Sigma_\nu [ (S_\nu - I_\nu) / \sigma_\nu ]^2 $ is a metric of spectral goodness-of-fit, where $S_\nu$ is the data spectrum, $I_\nu$ is the model spectrum, and $\sigma_\nu$ is the spectral noise estimate. However, note that this is not the same definition as the more commonly used `reduced chi-square' metric: $\chi^2/n$, where $n$ is the number of spectral points minus the number of degrees of freedom (model parameters). In this case therefore a good fit occurs when $\chi^2 \simeq n$ (rather than $\chi^2/n \simeq1$). We then define $\Delta \chi^2$ as the improvement to $\chi^2$ for various model trial abundances: $\Delta \chi^2 = \chi_{q}{^2} - \chi{_{0}}{^2}$, where $\chi{_{0}}{^2}$ denotes the best-fit model in absence of the trial gas, and $\chi_q^2$ is the same metric when an amount $q$ of the trial gas is present in the model \citep{teanby09a,nixon10b, nixon13b,teanby18}. An improved fit results in a $\Delta \chi^2$ that decreases below zero, and worsening fit results in $\Delta \chi^2$ that increases above zero. 

\begin{figure}[ht!]
\begin{centering}
\includegraphics[width=8cm]{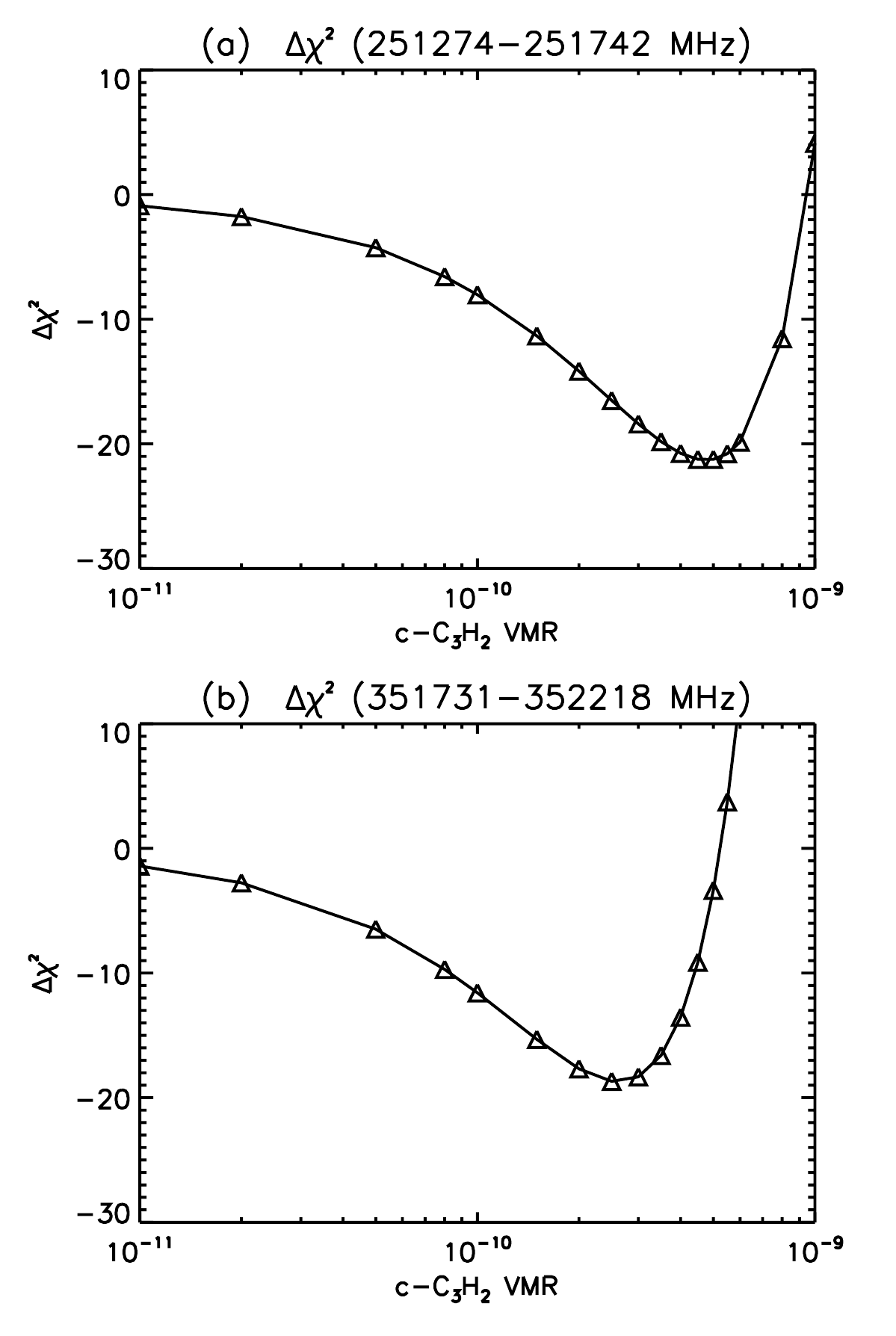}
\end{centering}
\caption{ 
Change in $\chi^2$ ($\Delta \chi^2$) for various trial abundances of c-\ccchh\ using a 350 km step model. 
(a) Band 6, showing a 4.6-$\sigma$ minimum at a volume mixing ratio (VMR) of 0.50 ppb. 
(a) Band 7, showing a 4.3-$\sigma$ minimum at a volume mixing ratio (VMR) of 0.25 ppb.
 \label{fig:c3h2-chisq}}
\end{figure}

Results are shown in Fig.~\ref{fig:c3h2-chisq}. A strong minimum is seen for a volume mixing ratio $q=0.5$~ppb in Band 6, with $\Delta \chi^2$ reaching -21.24 indicating a $\sqrt{21.24}$ =4.6~$\sigma$ significance to the result. For Band 7, a minimum is reached at mole fraction $q=0.25$~ppb with $\Delta \chi^2 = $~-18.69 (4.3~$\sigma$). Both results are  significant, although the mixing ratio determined in each case is somewhat different (but consistent within error bars, as shown later in Section~\ref{sect:results}). The combined significance of the detection is 6.3~$\sigma$.

Retrieved abundances for c-\ccchh\ with various profiles are described in Section~\ref{sect:results}.

\subsection{Spectral windows 2 and 3: Search for pyridine and pyrimidine}

\begin{figure}
\includegraphics[width=17cm]{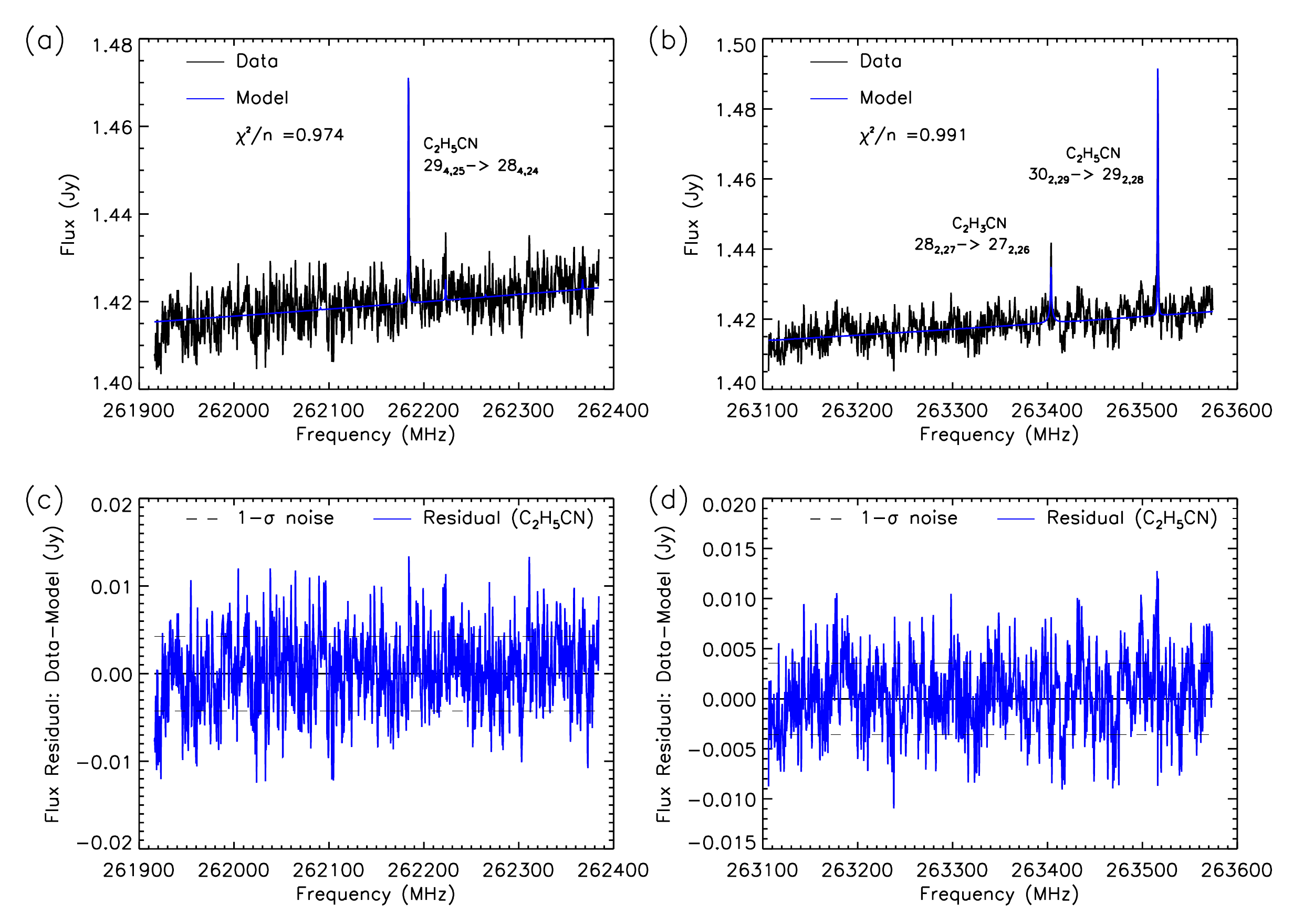}
\caption{ 
Best fit models for Spw 2 and 3. In Spw 2 (panel (a), and residual in panel (c)) we see a single strong line of \etcyn\ at 262183.8  ($29_{4,25} \rightarrow 28_{4,24}$), while in Spw 3 (panel (b), and residual in panel (d)) we detect \vycyn\ at 263403.7 MHz ($28_{2, 27} \rightarrow 27_{2, 26}$) ,  and \etcyn\ at 263516.2 MHz ($30_{2,29} \rightarrow 29_{2,28}$). 
Frequencies have been corrected to rest velocity frame.
}
 \label{fig:win23resid}
\end{figure}

Fitting for Spw 2 \& 3 was accomplished by initially using the retrieved temperature profile from Spw 4, and also scaling a 250-km step model for \etcyn\ and 300=km step model for \vycyn . In addition, we included HCN which contributed a continuum slope in these windows due to the wings of the strong 3$\rightarrow 2$ line at 265886~MHz whose line center lies outside the bandpass. Then, 300-km step model profiles models for c-\pyrid\ (Spw 2) and c-\pyrim\ (Spw 3) were introduced, but resulted in no significant improvement to the fit as measured by a reduced $\chi^2$ test. Instead, upper limits for c-\pyrid\ and c-\pyrim\ were determined instead (see Section~\ref{sect:results}.) The final fit for these spectral windows is shown in Fig.~\ref{fig:win23resid}.

\section{Results} 
\label{sect:results}

\subsection{Retrieval errors}

The propagation of errors in the retrieval process follows the formalism described in \citet{irwin08}, and further elaborated in Section 3.5 of \citet{nixon08a} (hereafter N08). This includes a combination of {\em a priori} and measurement error (Eq.~1 of N08), with the error from the earlier temperature propagated as additional measurement error (Eq.~2 of N08). In addition, we needed to make a further error allowance for apodization, which reduces independent information in the spectrum. Due to the Hanning apodization applied during the Fourier Transform, neighboring spectral channels become correlated, and the signal-to-noise in our retrieval will be overestimated by a factor equal to the square root of the number of channels per resolution element - two channels per resolution element for Hanning apodization. At the same time, there is a small gain of 1.095 from averaging information across two successive correlated channels{\footnote{ALMA technical notes: https://help.almascience.org/index.php?/Knowledgebase/Article/View/29}, so the final error bars are increased by a factor $\sqrt{2}/1.095 = 1.291$. This factor has also been applied to correspondingly reduce the detection significances ($\sigma$ levels) throughout the paper.

\subsection{Cyclopropenylidene}

We initially fitted the c-\ccchh\ emissions with a step function model, where the gas abundance was zero below a `step' altitude and a uniform value above. The overall profile was then scaled to achieve a best fit. The effect of changing the altitude of the step was also explored, since lower steps increased pressure broadening of the lines that became greater than the observed line widths.

We also investigated a more realistic gas profile, with an abundance decreasing downwards to a condensation altitude, by using a four-parameter gradient model. This model was defined by two $(p,q)$ (pressure, mixing ratio) co-ordinates defining a straight-line, logarithmically decreasing VMR between $(p_u,q_u)$ (upper point) and $(p_l,q_l)$ (lower point). Above $p_u$ the VMR was assumed constant at $q_u$ and below $p_l$ the VMR dropped to zero. The upper pressure level was set to be $p_u = (5.0 \pm 2.0) \times 10^{-11}$ bar, or approximately 1100 km, the altitude of the INMS measurements of the \ccchh H$^+$ (protonated) ion. The initial value for the abundance at this altitude was set to be 
 $q_u = (3.4 \pm 1.0) \times 10^{-6} $ in line with the INMS ion measurements \citep{vuitton07}.
The initial value for the lower point was set to be: 
$p_l = (1.0 \pm 0.5) \times 10^{-4}$ bar, $q_l = (2.0 \pm 1.9) \times 10^{-9} $, a pressure level corresponding to approximately 300 km, and allowing a lenient variation of abundance. 

Scaled step function solutions for c-\ccchh\ from Window 1 (Band 6) and Window 6 (Band 7) are shown in Fig.~\ref{fig:profs}, along with best fit gradient model profiles. Numerical results are given in Table~\ref{tab:ret}.
Retrievals for \cpld\ showed low sensitivity to the altitude of the step, with a weak minimum at 300--400 km. 
The resulting abundances and columns were slightly different in 2016 and 2017.
For a step function of 350 km we obtained a VMR of 0.50 $\pm$ 0.14 ppb and column abundance of 3.5$\times 10^{12}$ cm$^{-2}$ in 2016, but somewhat lower VMR (0.28 $\pm$ 0.08) and column abundance (1.5$\times 10^{12}$) in 2017. 
This implies either (a) that the global abundance had decreased from 2016 to 2017, or, (b) if the real abundance was constant, then a mean value of 0.33 $\pm$ 0.07 ppb for the 350 km step profile.

\begin{figure}[ht!]
\includegraphics[width=18cm]{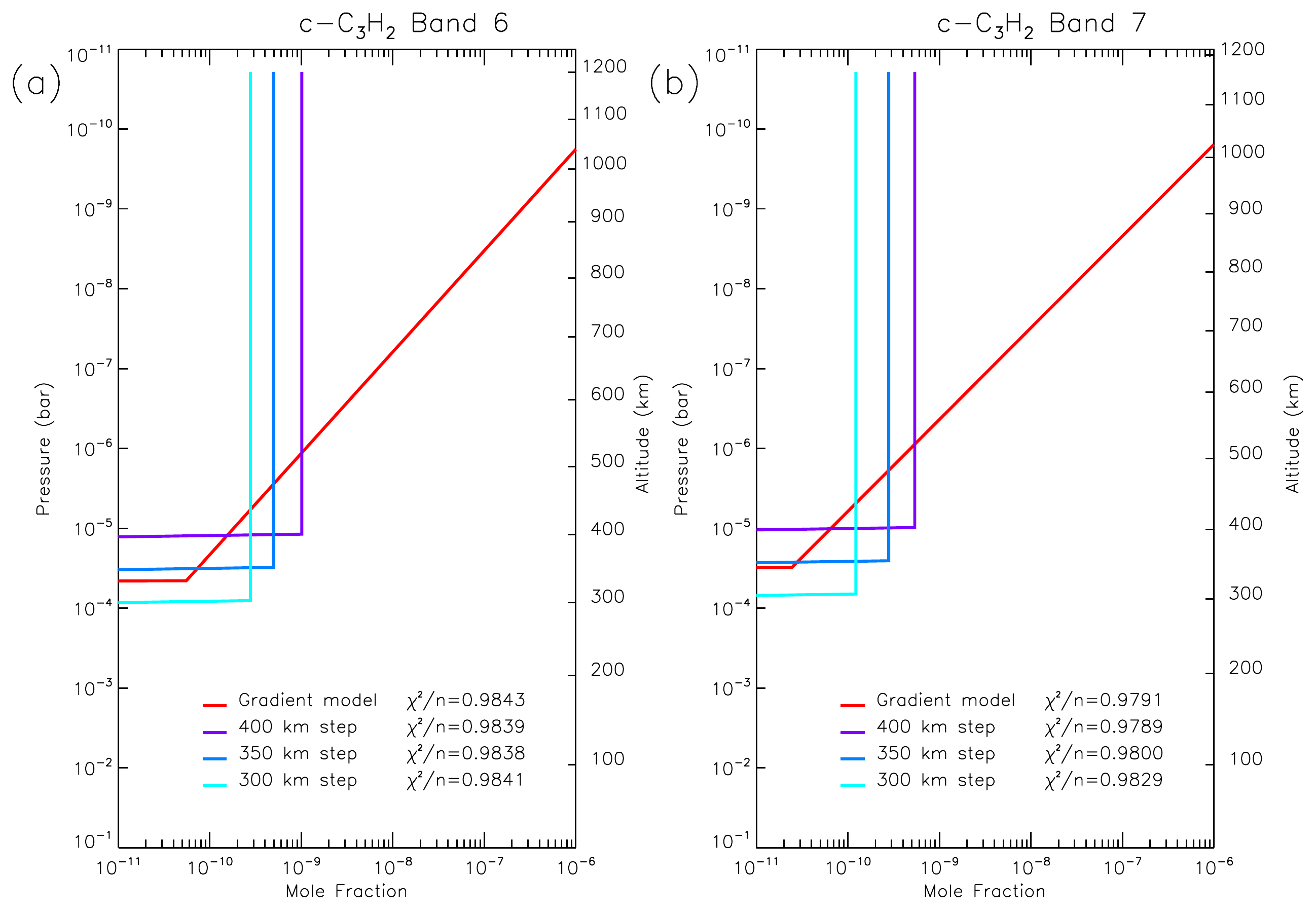}
\caption{ 
Retrieved profiles of c-\ccchh\ for different models. (a) Band 6 data, 2016. (b) Band 7 data, 2017.
 \label{fig:profs}}
\end{figure}

\begin{table}[ht]
\caption{Retrieved column abundances and volume mixing ratios (VMRs) at 600 km for different c-\ccchh\  models. }
\label{tab:ret}
\begin{tabular}{lllccc}
& & & & & \\
\hline
Band & Species & Model & $\chi^2$/n & VMR & Col. Abund. \\
&  & & & (ppb @ 600 km) & (molecule cm$^{-2}$)  \\
\hline
& & & & & \\
6 & c-C$_3$H$_2$ &  Gradient model & 0.9843 &   3.788 &  2.649$\times 10^{12}$ \\
6 & c-C$_3$H$_2$ &     400 km step & 0.9839 &   1.012$\pm$  0.386 &   2.824$\times 10^{12}$ \\
6 & c-C$_3$H$_2$ &     350 km step & 0.9838 &   0.495$\pm$  0.142 &   3.487$\times 10^{12}$ \\
6 & c-C$_3$H$_2$ &     300 km step & 0.9841 &   0.278$\pm$  0.054 &   4.875$\times 10^{12}$ \\
& & & & & \\
7 & c-C$_3$H$_2$ &  Gradient model & 0.9791 &   1.867 &  1.197$\times 10^{12}$ \\
7 & c-C$_3$H$_2$ &     400 km step & 0.9789 &   0.537$\pm$  0.223 &   1.175$\times 10^{12}$ \\
7 & c-C$_3$H$_2$ &     350 km step & 0.9800 &   0.279$\pm$  0.084 &   1.541$\times 10^{12}$ \\
7 & c-C$_3$H$_2$ &     300 km step & 0.9829 &   0.122$\pm$  0.031 &   1.702$\times 10^{12}$ \\
\hline
& & & & & \\
\end{tabular}
\end{table}

Retrieved parameters for the gradient model retrievals in 2016 and 2017 are shown in Table \ref{tab:grad}, along with parameters for a weighted mean profile of both years.

\begin{table}
\caption{Retrieved parameters for c-\ccchh\ gradient model fits \label{tab:grad}}
\begin{tabular}{ccccc}
\hline
 & $p_u$ (bar) & $q_u$ & $p_l$ (bar) & $q_l$ \\
\hline
{\em Band 6} & & & \\
{\em a priori} & (5.0$\pm$2.0) $\times 10^{-11}$ &  (3.4$\pm$1.0) $\times 10^{ -6}$ & (1.0$\pm$0.5) $\times 10^{ -4}$ & (2.0$\pm$1.9) $\times 10^{ -9}$ \\ 
Retrieved & (4.4$\pm$2.2) $\times 10^{-11}$ &  (3.1$\pm$1.2) $\times 10^{ -6}$ & (4.7$\pm$2.9) $\times 10^{ -5}$ & (5.4$\pm$3.7) $\times 10^{-11}$ \\ 
& & & & \\
{\em Band 7} & & & \\
{\em a priori} & (5.0$\pm$2.0) $\times 10^{-11}$ &  (3.4$\pm$1.0) $\times 10^{ -6}$ & (1.0$\pm$0.5) $\times 10^{ -4}$ & (2.0$\pm$1.9) $\times 10^{ -9}$ \\ 
Retrieved & (4.3$\pm$2.2) $\times 10^{-11}$ &  (3.1$\pm$1.2) $\times 10^{ -6}$ & (3.8$\pm$2.3) $\times 10^{ -5}$ & (2.1$\pm$1.7) $\times 10^{-11}$ \\ 
& & & & \\
{\em Band 6 \& 7} & & & \\
Combined & (4.3$\pm$1.6) $\times 10^{-11}$ &  (3.1$\pm$0.8) $\times 10^{ -6}$ & (4.1$\pm$1.8) $\times 10^{ -5}$ & (2.6$\pm$1.5) $\times 10^{-11}$ \\ 
\hline
\end{tabular}
\end{table}

\subsection{Pyridine and pyrimidine}

Upper limits for c-\pyrid\ and c-\pyrim\ were determined using the $\Delta \chi^2$ method outlined for c-\ccchh\ in Section \ref{sect:win16}. In this case, the 1, 2 and 3-$\sigma$ upper limits are indicated at the trial abundances where the $\Delta \chi^2$ reaches +1, +4, and +9 respectively \citep{nixon12b}. Results are shown in Fig.~\ref{fig:panh_ulimits} and Table \ref{tab:panhs}. A shallow minimum was detected for c-\pyrid , however the spectrum does not show obvious emissions consistent with expected spectral lines, therefore we believe this to be likely due to random spectral noise (although worthy of a more sensitive follow-up observation to be sure).

\begin{table}
\caption{Upper limits for undetected nitrogen heterocycle molecules in Titan's atmosphere \label{tab:panhs}}
\begin{tabular}{lllllll}
\hline
Name & p ($\mu$bar) & Freq. (MHz) & 
NEF$^{\dagger}$ (mJy) &  1-$\sigma$ VMR$^{\ddagger}$ & 2-$\sigma$ VMR$^{\ddagger}$ & 3-$\sigma$ VMR$^{\ddagger}$ \\
c-\pyrim\ &  0.020 & 262143 & 0.34 &  0.663 &  0.854 &  1.042 \\
c-\pyrid\  &  0.020 & 263331 & 0.29 &  1.046 &  1.153 &  1.356 \\  
\hline
 & & & & & & \\
 \end{tabular}
 \newline
 $^\dagger$Noise Equivalent Flux. $^{\ddagger}$Volume Mixing Ratio (mole fraction) in ppb.
\end{table}

\begin{figure}[ht!]
\includegraphics[width=18cm]{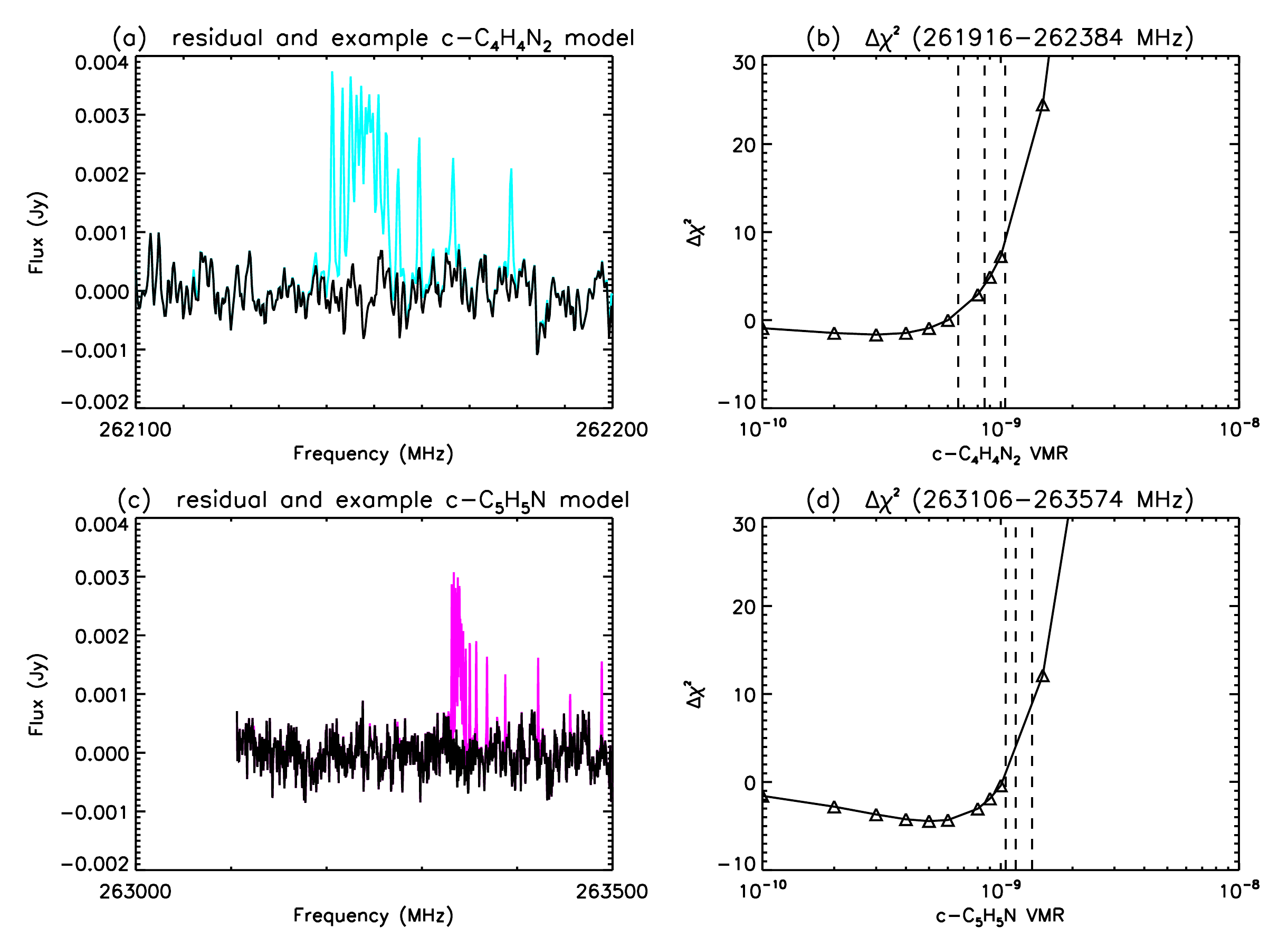}
\caption{ 
Upper limit determination for nitrogen heterocycle molecules. 
(a) Data (black) and example spectrum (cyan) for c-\pyrim\ showing missing spectral band location. (b) 
$\Delta \chi^2$ curve for various trial abundances for c-\pyrim . 
(c) Data (black) and example spectrum (magenta) for c-\pyrid\ showing missing spectral band location. (d) $\Delta \chi^2$ curve for various trial abundances for c-\pyrid . Vertical dashed lines on (b) and (d) indicate the 1, 2, and 3-$\sigma$ upper abundance limits at $\Delta \chi^2 = +1, +4, +9$.
\label{fig:panh_ulimits}}
\end{figure}

\section{Discussion} 
\label{sect:disc}

\subsection{Cyclopropenylidene}

The molecule \cpld\ (c-\ccchh ) was discovered in the interstellar medium (ISM) by \citet{thaddeus85} through extensive laboratory and theoretical analysis to unearth the origin of several prominent, but previously unidentified lines seen on radio astronomical spectra. Following this discovery, the molecule has been found to be ubiquitous in the galaxy \citep{fosse01} and easily detectable due to the relatively large dipole of 3.43(2) D \citep{kanata87} caused by the unpaired electrons on the bivalent carbon atom. In addition, c-\ccchh\ is a light molecule with a small partition function, which also works in favor of detection. One of its linear isomers, \pdld\ (H$_2$CCC, see Fig.~\ref{fig:c3s}) has since been detected in the ISM \citep{cernicharo91} while \ppld\ (HCCCH) has not been observed. Note that \pdld\ is higher in energy than \cpld , and therefore metastable, so that the observed ratio of ten or more for c-\ccchh /H$_2$CCC is expected.

The Cassini INMS instrument measured peaks at $m/z$ 38 and 39 in samples of Titan's upper atmosphere that were attributed to the presence of C$_3$H$_2^+$ and various isomers of C$_3$H$_3^+$  \citep{vuitton06a, vuitton07}.  Although the molecular structure was not directly measurable, modeling of the mass spectrum implied ion number densities of 0.0016 cm$^{-3}$ (C$_3$H$_2^+$), 34 cm$^{-3}$ (c-C$_3$H$_3^+$) and 1.6 cm$^{-3}$ (l-C$_3$H$_3^+$) respectively. Determining the ratio of l-C$_3$H$_3^+$/c-C$_3$H$_3^+$ was deemed to be of major importance by \citet{vuitton07} (and the subject of laboratory investigation), since the linear propargyl ion is able to react to form heavier species, including possibly benzene \citep{wilson04}, while the \cpld\ ion is essentially a terminal species, leading to c-\ccchh .

\begin{figure}[ht!]
\includegraphics[width=18cm]{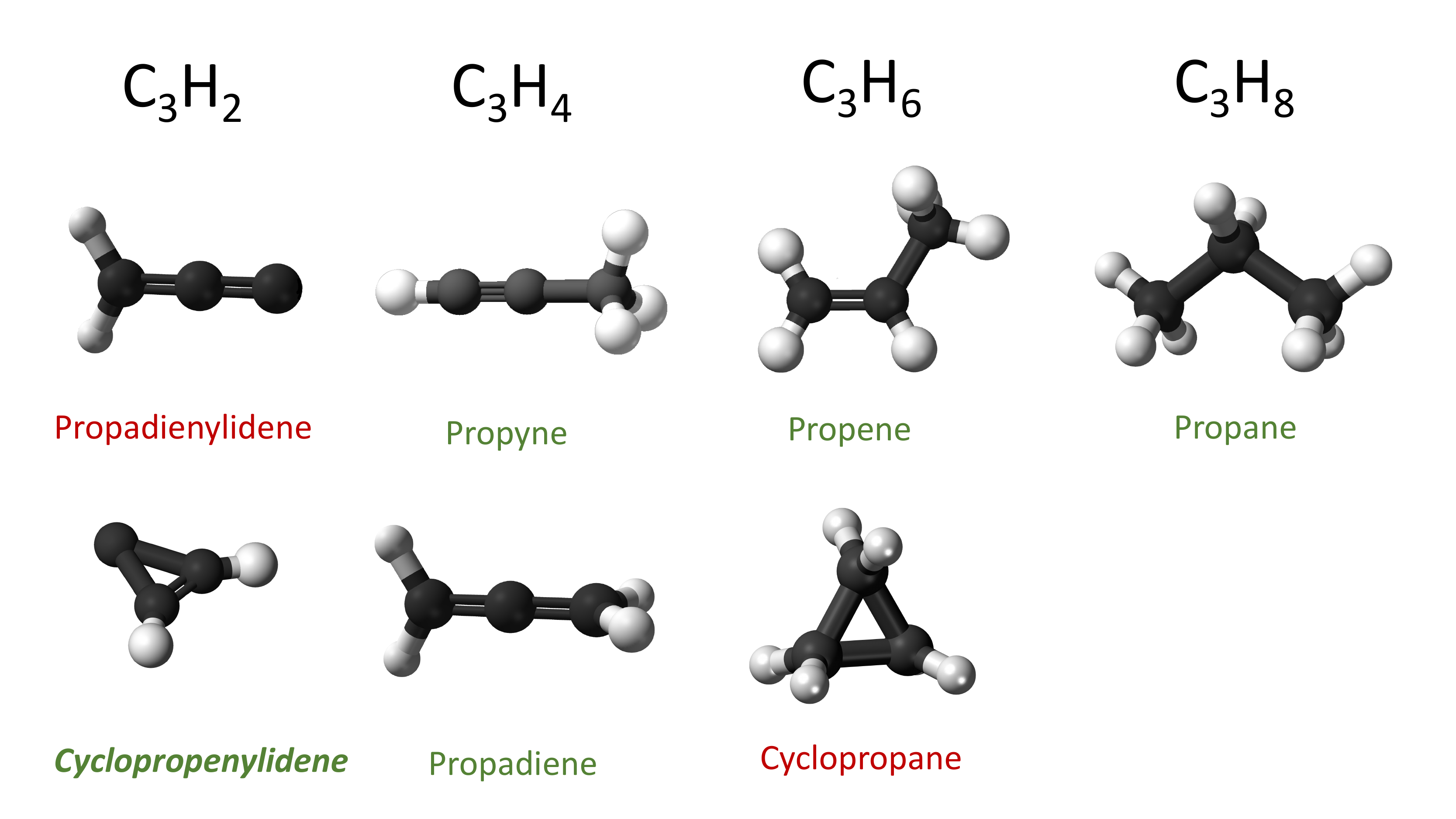}
\caption{Structures of C$_3$H$_x$ molecules. Green lettering = detected in Titan's atmosphere. Red lettering = undetected. Molecule graphics: wikimedia commons.
  \label{fig:c3s}}
\end{figure}

Various mechanisms have been proposed for the formation of c-\ccchh\ and it is not clear at present which mechanisms are the most important.
In the original work of \citet{thaddeus85} on the ISM detection, cyclopropenylidene was produced by dissociative electron recombination of the cyclopropenylium cation, c-C$_3$H$_3^+$: 

\begin{equation}
{\rm c{-}C_3 H_3{^+} + e^-  \rightarrow c{-}C_3 H_2 + H \label{eq:recombination}}
\end{equation}

\noindent while c-C$_3$H$_3^+$ is produced from \acet\ in two steps. First the fast ion-molecule reaction:

\begin{equation}
{\rm C^+ + C_2H_2  \rightarrow c{-}C_3H^+ + H }
\end{equation}

\noindent followed by the slower radiative association (hydrogenation):

\begin{equation}
{\rm c{-}C_3H^+ + H_2 \rightarrow c{-}C_3H_3^+    \label{eq:hydrogenation}}
\end{equation}

Alternatively the C$_3$H$_3^+$ ion has been proposed to be produced from acetylene via many other possible ion molecule reactions by \citet{vuitton19}, for example:

\begin{equation}
{\rm CH_3^+ + C_2H_2 \rightarrow C_3H_3^+ + H_2 }
\end{equation}
\begin{equation}
{\rm C_2H_5^+ + C_2H_2 \rightarrow C_3H_3^+ + CH_4  }
\end{equation}

 \citet{walch95, guadagnini98} investigated the reactions of CH($X\:  ^2\Pi$)  (methylidyne) with \acet , predicting that various isomers of both C$_3$H$_3$ and C$_3$H$_2$ can result. From this point, several outcomes are possible: the products can stabilize into a less-reactive species, such as c-\ccchh , or else can undergo further reactions to form heavier hydrocarbons. In particular, it was noted that both C$_3$H$_3$ and C$_3$H$_2$ can dimerize, forming benzene (\benzene ) and $para$-benzene (C$_6$H$_4$) respectively, and therefore C$_3$H$_3$ and C$_3$H$_2$ are important stepping stones to polycyclic aromatic hydrocarbons (PAHs). 

The work of \citet{canosa97} further clarified  pathways to formation of \ccchh\ from reactions of the methylidine radical (CH) with unsaturated C$_2$H$_x$ hydrocarbons, such as:

\begin{equation}
{\rm CH + C_2H_2 \rightarrow C_3H_2 + H , or }
\end{equation}
\begin{equation}
{\rm CH + C_2H_4 \rightarrow C_3H_3 + H_2  }
\end{equation}

\noindent
In the above reactions, CH is envisaged to add to the carbon-carbon double or triple bond. C$_3$H$_3$ can be converted to C$_3$H$_2$ by hydrogen loss through photodissociation \citep[e.g.][]{hebrard13}:

\begin{equation}
{\rm C_3H_3 + h\nu \rightarrow C_3H_2 + H }
\end{equation}

\noindent 
The C$_3$H radical may also result from the methylidine insertion reactions, which can lead to \ccchh\ via several steps, first charge transfer:

\begin{equation}
{\rm H^+ + C_3H \rightarrow C_3H^+ + H}
\end{equation}

\noindent followed by hydrogenation (\ref{eq:hydrogenation}) and then dissociative recombination (\ref{eq:recombination}) as before.
Subsequently, \citet{canosa07} showed that C$_2$ reactions may also be important, e.g,:

\begin{equation}
{\rm C_2 +C_3H_8 \rightarrow C_3H_2 + C_2H_6 }
\end{equation}

\noindent
as used in the photochemical model of \citet{krasnopolsky09}. The branching ratios between aliphatic and aromatic pathways in many of these reactions, especially at low temperatures, are important and often poorly known.

In Fig.~\ref{fig:photochem} we compare our retrieved gradient models to photochemical model predictions of \citet{hebrard13} and \citet{vuitton19}. In fact, the models arrive at a column abundance rather similar to our retrieved amount of $\sim$10$^{12}$~cm$^{-2}$ . It is difficult to pronounce whether the differences in the vertical profile shape are significant or not, since we have very little constraint on this at present.

\begin{figure}[ht!]
\includegraphics[width=18cm]{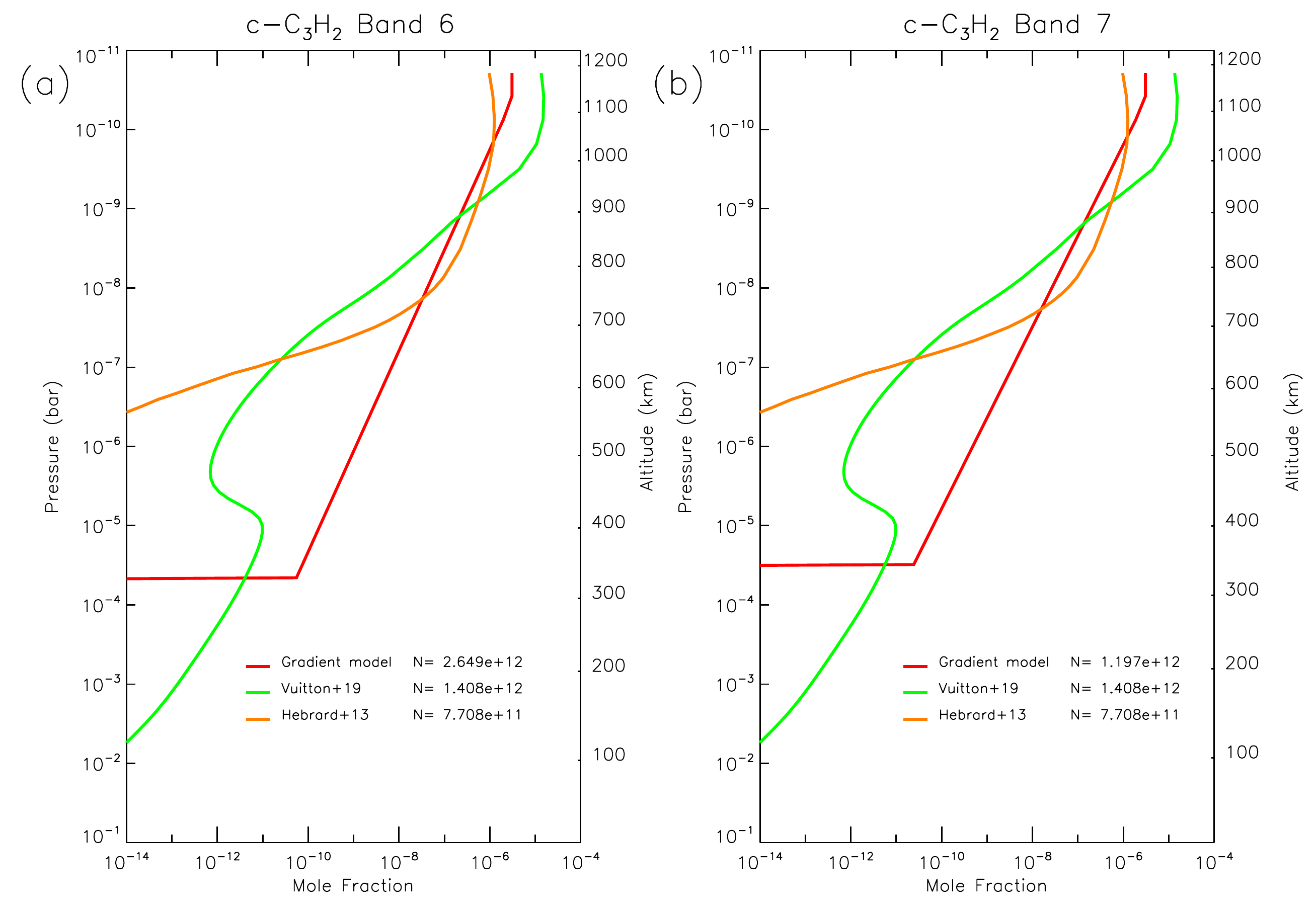}
\caption{ 
Retrieved gas abundance profiles for c-\ccchh\ in Band 6 and Band 7 using the gradient model compared to published photochemical models. Column abundances (N) are in molecule cm$^{-2}$.
\label{fig:photochem}}
\end{figure}

\subsection{Pyridine and pyrimidine}

The astrobiologically important species pyridine (c-\pyrid ) and pyrimidine (c-\pyrim ) are nitrogen-containing heterocyclic ring molecules resembling a benzene ring with either one or two of the C-H members replaced by a nitrogen atom. Pyrimidine in particular is of significant biological importance since it forms the backbone ring structure of several key biological molecules - specifically the nucleobases uracil (in RNA), cytosine (in RNA and DNA) and thymine (in DNA). These molecules can potentially be formed from pyrimidine after the chemical substitution of functional groups (-NH$_2$, -CH$_3$ and =O) in place of hydrogen, as indicated in Fig.~\ref{fig:panhs}. Indeed, laboratory experiments \citep{nuevo14} have shown that UV irradiation of pyrimidine in the presence of \water , CH$_4$, CH$_3$OH and NH$_3$ can form uracil and cytosine - but not the more complex thymine - a possible clue as to why thymine appears only in DNA but not RNA, and further evidence that RNA may have preceded DNA. Similar processes may be taking place in space, including the atmosphere of Titan. Indeed, laboratory simulations of Titan's atmosphere, using multiple experimental techniques such as GC-MS (gas chromatograph mass spectroscopy), pyrolysis mass spectroscopy, Raman and reflectance spectroscopy etc have been successful in positively identifying the nitrogen heterocycles \citep{khare84b, ehrenfreund95}.

\begin{figure}[ht]
\includegraphics[width=17cm]{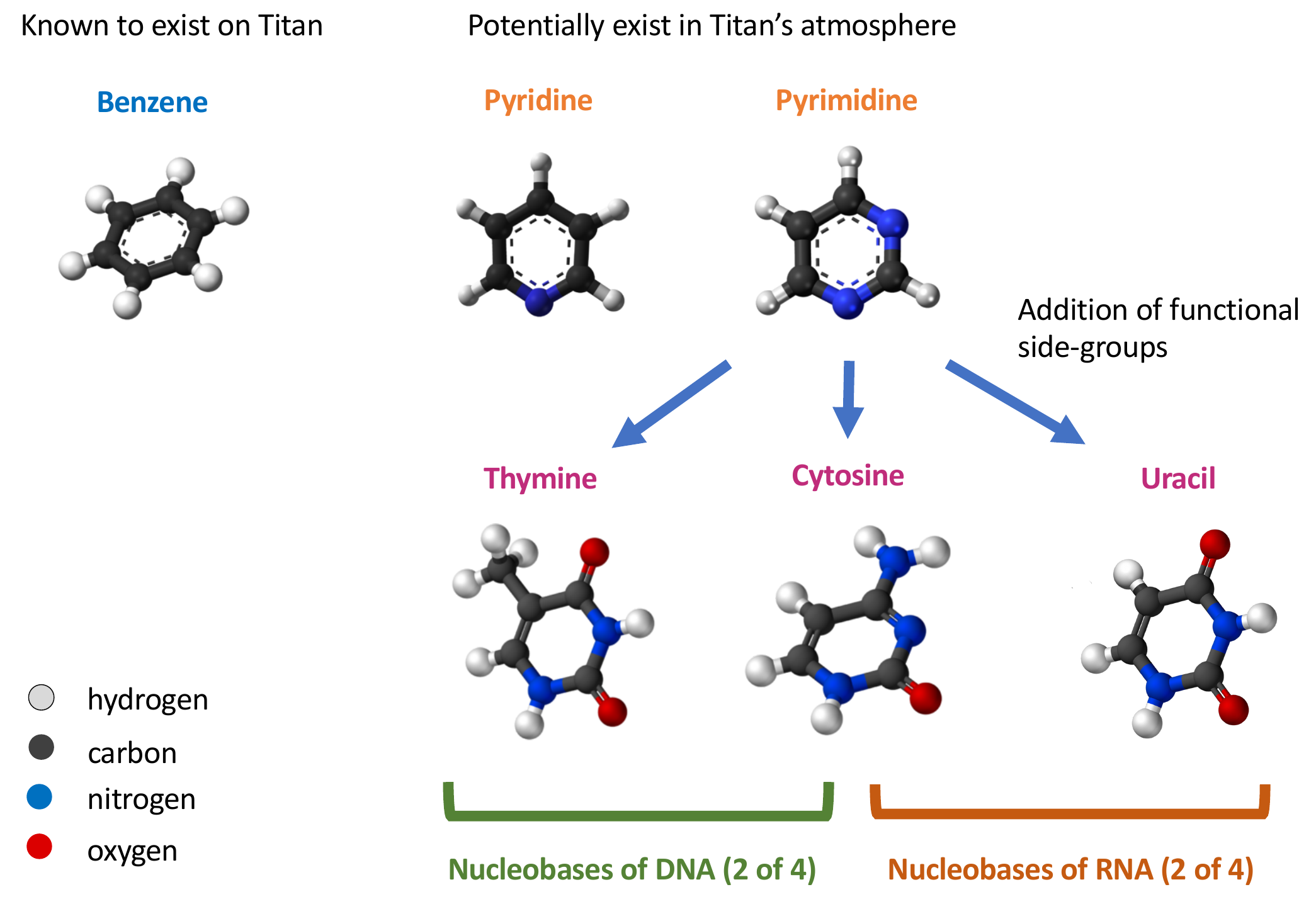}
\caption{Importance of  nitrogen heterocycle molecules such as pyridine and pyrimidine to astrobiology: detection of such species would indicate a possible route to creation of nucleobases of DNA and RNA. Individual molecule graphics from wikimedia commons.
\label{fig:panhs}}
\end{figure}

To date, neither pyridine nor pyrimidine have been detected in astrophysical sources, despite searches in molecular clouds \citep{simon73,kuan03, kuan04, cordiner17, mcguire18} and in the envelopes of evolved stars \citep{charnley05}, although pyridine and quinoline (2-membered N-heterocycle rings) derivatives have been found in meteorite samples \citep[e.g.][]{stoks82, martins18}.  \citet{peeters05} have shown that these molecules are relatively unstable against UV irradiation compared to benzene, but could survive for 10--100s of years in dense clouds where UV flux is attenuated, and therefore potentially in Titan's thick atmosphere. 

The potential presence of the nitrogen heterocyclic molecules pyridine and pyrimidine in Titan's atmosphere may be inferred from the detection of \pyrid H$^+$ and \pyrim H$^+$ ions in Cassini mass spectra \citep{vuitton07} at $m/z$ 80 and 81 (seen in their Fig.~2). As with the hydrocarbons, the elucidation of structure from the mass spectra alone is not possible, therefore for example protonated forms of branched acyclic molecules such as penta-2,4-dienenitrile or 2-methylene-3-butenenitrile could be responsible for the mass 80 peak instead. 

Formation pathways for the N-heterocycles are currently quite uncertain. For example, \citet{fondren07} suggest that efficient ion-molecule association reactions with HCN could form pyridine and pyrimidine from smaller ions:

\begin{equation}
{\rm C_4H_4^+ + HCN \rightarrow  C_5H_5N^+ + h\nu}
\end{equation}
\begin{equation}
{\rm C_3H_3N^+ + HCN \rightarrow  C_4H_4N_2^+ + h\nu}
\end{equation}

\noindent
A more exotic mechanism for the formation of pyridine through ring expansion of pyrrole by methylidyne has been observed in the gas phase by \citet{soorkia10}:

\begin{equation}
{\rm CH + C_4H_5N \rightarrow C_5H_5N + H }
\end{equation}

\noindent
More recently, \citet{balucani19} has investigated a pathway to pyridine that begins with an attack on \benzene\ by N($^2$D), leading to a chain of unstable intermediate products that may decay to c-\pyrid .

The relative importance of these various channels is highly uncertain at the present time, leading to difficulties in incorporating these molecules into photochemical models. For example \citet{krasnopolsky09} included just one hypothetical formation pathway for pyridine by the radical-molecule reaction:

\begin{equation}
{\rm C_3N + C_2H_6 \rightarrow C_5H_5N + H }
\end{equation}

\noindent
while in \citep{loison15} only the aliphatic isomer C$_2$H$_5$C$_3$N is discussed.

Our analysis indicates 2-$\sigma$ upper limits of $\sim 1.15$ ppb and $\sim 0.85$ ppb for c-\pyrid\ and c-\pyrim\ respectively (constant profile above 300 km), which may in future be used to place some constraints on photochemical models as these become more sophisticated and add more detailed treatment of cyclic molecule formation.

\section{Conclusions} 
\label{sect:conc}

We report the first detection of c-\ccchh\ (\cpld ) on Titan in two datasets: Band 6 spectra from 2016 and Band 7 data from 2017, detecting at least 2 emissions in each case. The derived abundances are 0.50 $\pm$ 0.14 ppb in 2016 and 0.28 $\pm$ 0.08 in 2017 for a 350-km step model, which are in agreement at the margins of their 1-$\sigma$ errors, or alternatively may indicate a real decrease in abundance. Derived column abundances are 3--5  $\times 10^{12}$~cm$^{-2}$ in 2016 and 1--2  $\times 10^{12}$~cm$^{-2}$ in 2017, in good agreement with photochemical models. This presence of \cpld\ is of substantial significance to Titan's atmospheric chemistry, since insertion reactions of methylidyne (CH) into \acet\ and other unsaturated hydrocarbons can lead to formation of \ccchh\ and C$_3$H$_3$ isomers. These in turn may be stepping stones to benzene and $para$-benzene, and larger aromatic PAH molecules. 

Following preliminary evidence from Cassini mass spectra, we also searched for the N-heterocyclic molecules pyridine and pyrimidine in Titan's atmosphere, with a null result. By modeling of ALMA spectra at 262--263 GHz we have determined 2-$\sigma$ upper limits of 1.15 and 0.85 ppb for c-\pyrid\ and c-\pyrim\ respectively. We have detected ground state lines of \vycyn\ and \etcyn\ as previously seen in Titan's atmosphere, and also vibrationally excited rotational transitions of \etcyn . The \etcyn\ emissions are well-fitted using a 250 km `step' model as noted by previous authors, and we find a best-fit abundance of 5.0 $\pm$ 0.1 ppb similar to previous work. Our modeling indicates that there is unlikely to be substantial amounts of \etcyn\ below 250~km, in contrast to existing photochemical models. 

The discovery of \cpld\ for the first time in a dense planetary atmosphere therefore opens up new directions for research in the chemistry of the reducing atmospheres of the outer planets, and especially PAH and haze formation.

\acknowledgements

C.A.N. and M.A.C. received support for this work through NASA's Solar System Observations (SSO) Program. C.A.N. and A.E.T. were also funded by NASA's Astrobiology Program. M.A.C. was supported by the National Science Foundation under Grant No. AST-1616306. PGJI and NAT are funded by the UK Science and Technology Facilities Council.

This paper makes use of the following ALMA data: ADS/JAO.ALMA\#2015.1.00423.S, 2015.1.00512.S, 2016.A.00014.S. ALMA is a partnership of ESO (representing its member states), NSF (USA) and NINS (Japan), together with NRC (Canada), MOST and ASIAA (Taiwan), and KASI (Republic of Korea), in cooperation with the Republic of Chile. The Joint ALMA Observatory is operated by ESO, AUI/NRAO and NAOJ. The National Radio Astronomy Observatory is a facility of the National Science Foundation operated under cooperative agreement by Associated Universities, Inc.

\appendix

\section{Propionitrile}
\label{sect:c2h5cn}

\subsection{Modeling}

In Spw 1, sufficiently strong lines of \etcyn\ were seen that there was noticeable pressure-induced line broadening, allowing an optimum altitude for a step function model to be determined.  Fig.~\ref{fig:c2h5cn} shows the effect of changing the step function altitude for \etcyn . 

\begin{figure}
\includegraphics[width=18cm]{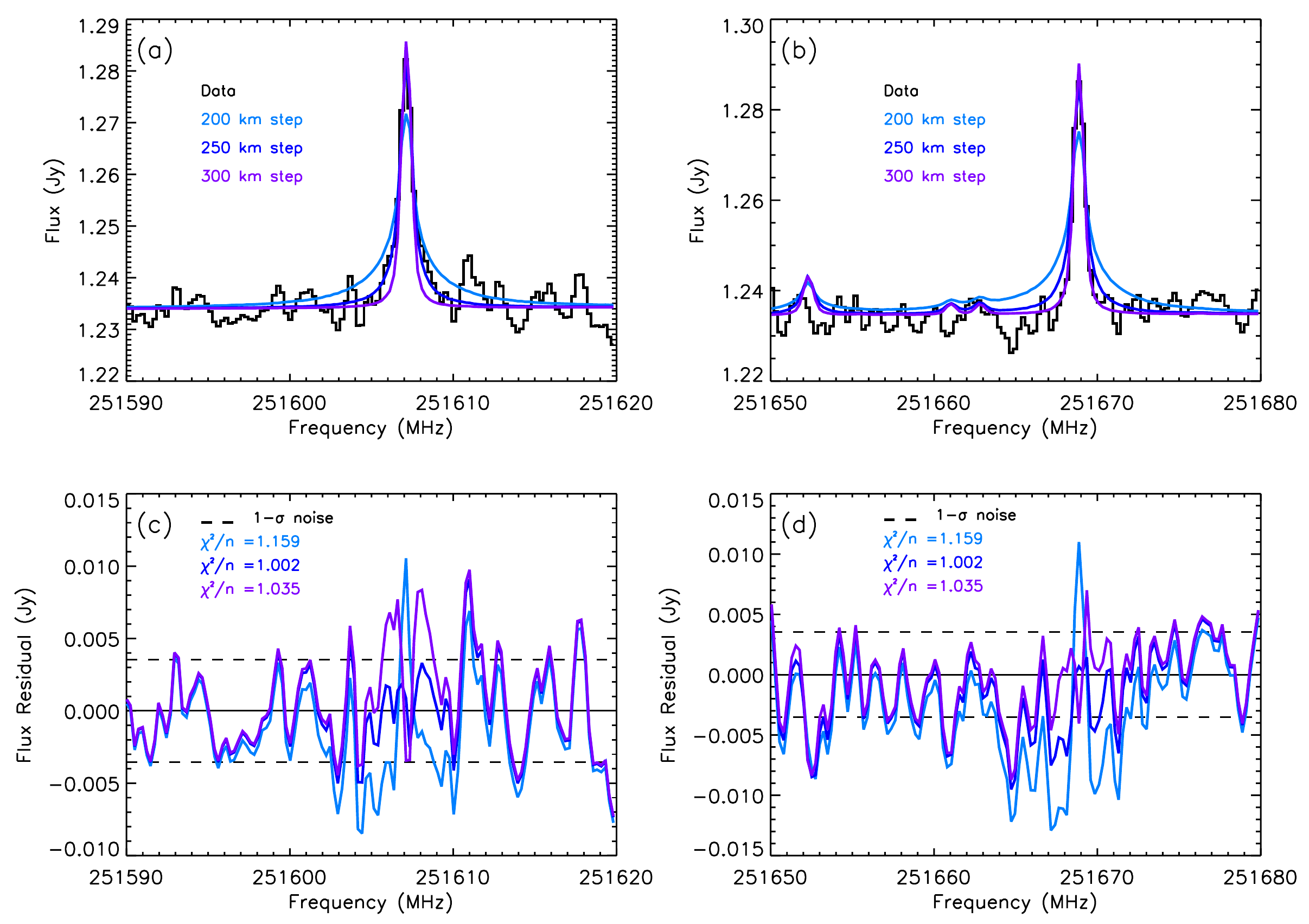}
\caption{  
Spectral fitting of strong \etcyn\ lines in Spw 1 using `step' models at different altitudes (panels (a) and (b)), and residual after fitting (panels (c) and (d)). 
Frequencies have been corrected to rest velocity frame.
\label{fig:c2h5cn}}
\end{figure}

The goodness of fit for all models is compared in Table~\ref{tab:c2h5cn}. The best-fit solution for \etcyn\ is a step function at 250 km with a uniform VMR of 5.0 $\pm$ 0.07 ppb and column abundance 2.2$\times 10^{14}$ cm$^{-2}$.  
A comparison of retrieved profiles can be seen in Fig.~\ref{fig:c2h5cn_profs}(a).

\begin{table}[ht]
\caption{Retrieved column abundances and volume mixing ratios (VMRs) at 600 km for different model types for \etcyn . }
\label{tab:c2h5cn}
\begin{tabular}{llccc}
& & & & \\
\hline
Species & Model & $\chi^2 /n$ & VMR  & Col. Abund \\
 & & & (ppb @ 600 km) & (molecule cm$^{-2}$)  \\
\hline
C$_2$H$_5$CN &  Gradient model & 1.051 &  31.957                   &   1.3475$\times 10^{14}$ \\
C$_2$H$_5$CN &     300 km step & 1.035 &   8.545$\pm$  0.252 &   1.5004$\times 10^{14}$ \\
C$_2$H$_5$CN &     250 km step & 1.002 &   5.040$\pm$  0.095 &   2.2378$\times 10^{14}$ \\
C$_2$H$_5$CN &     200 km step & 1.159 &   2.749$\pm$  0.030 &   3.5703$\times 10^{14}$ \\
\hline
& & & & \\
\end{tabular}
\end{table}

\begin{figure}
\includegraphics[width=18cm]{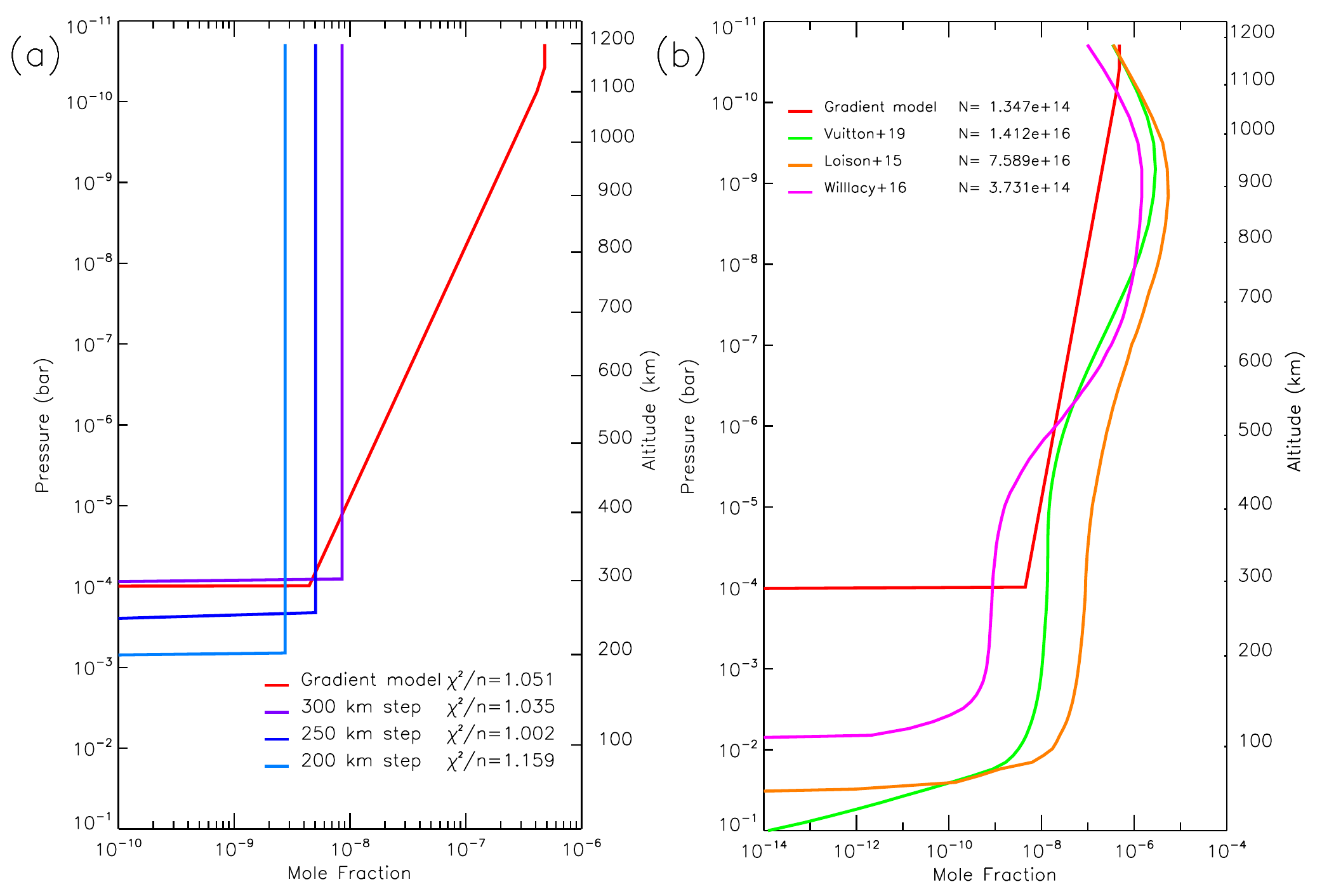}
\caption{  
(a) Retrieved model profiles for \etcyn . (b) Comparison between gradient model for \etcyn\ and published photochemical model profiles. 
\label{fig:c2h5cn_profs}}
\end{figure}

A gradient model was also tested, using similar initial conditions to those used for c-\ccchh , as described in Section~\ref{sect:win16}, except that the initial abundance at the 1100~km altitude was set to $q_u = (5.0 \pm 1.0) \times 10^{-7}$ in line with the INMS measurement of \etcyn H$^+$. Initial and retrieved parameters for the \etcyn\ gradient model are given in Table~\ref{tab:gradc2h5cn}.

\begin{table}
\caption{Retrieved parameters for gradient model fits \label{tab:gradc2h5cn}}
\begin{tabular}{cccccc}
\hline
Gas & & $p_u$ (bar) & $q_u$ & $p_l$ (bar) & $q_l$ \\
\hline
\etcyn\ & {\em a priori} & $(5.0 \pm 2.0) \times 10^{-11}$ & $(5.0 \pm 1.0) \times 10^{-7}$ & $(1.0 \pm 0.5) \times 10^{-4}$ & $(2.0 \pm 1.9) \times 10^{-9}$ \\
\etcyn\ & Retrieved      & $(4.7 \pm 2.4) \times 10^{-11}$ & $(4.8 \pm 1.3) \times 10^{-7}$ & $(1.1 \pm 0.6) \times 10^{-4}$ & $(4.3 \pm 0.9) \times 10^{-9}$ \\
\hline
\end{tabular}
\end{table}

A list of the ground-state and vibrationally excited lines detected in Spw 1 is given in Table~\ref{tab:lines}.

\begin{table}[h]
\caption{Lines of \etcyn\ detected in Spw 1 }
\begin{singlespace}
\small
\begin{tabular}{lcccc}
& & & & \\
\hline
Species & Freq. (MHz) & Transition$^a$  & $v^b$ & $E_u$ (K) \\
\hline
C$_2$H$_5$CN & 251271.3 & $30_{ 6,24}$-$30_{ 5 ,25}$ & 0 & 240 \\
C$_2$H$_5$CN & 251278.7 & $28_{18,}$-$27_{18,}$  & 0 & 533 \\
C$_2$H$_5$CN & 251284.2 & $28_{16,}$-$27_{16,}$ & 2 & 776 \\
C$_2$H$_5$CN & 251289.1 & $28_{12,}$-$27_{12,}$ & 1 & 624 \\
C$_2$H$_5$CN & 251297.1 & $28_{11,}$-$27_{11,}$ & 1 & 600  \\
C$_2$H$_5$CN & 251302.3 & $28_{13,}$-$27_{13,}$ & 1 & 651  \\
C$_2$H$_5$CN & 251331.4 & $28_{14,}$-$27_{14,}$ & 1 & 680 \\
C$_2$H$_5$CN & 251335.9 & $28_{10,}$-$27_{10,}$ & 1 & 577  \\
C$_2$H$_5$CN & 251365.8 & $28_{19, }$-$27_{19, }$ & 0 & 573  \\
C$_2$H$_5$CN & 251373.2 & $28_{15,}$-$27_{15,}$ & 1 & 712  \\
C$_2$H$_5$CN & 251404.3 & $28_{ 6,23}$-$27_{ 6,22}$  & 3 & 751 \\
 C$_2$H$_5$CN & 251409.4 & $28_{ 6,22}$-$27_{ 6,21}$ & 3 & 751  \\
C$_2$H$_5$CN & 251419.7 & $28_{ 9,}$-$27_{ 9,}$ & 1 & 557  \\
C$_2$H$_5$CN & 251425.6 & $28_{16,}$-$27_{16,}$ & 1 & 744  \\
C$_2$H$_5$CN & 251459.0 & $28_{20, }$-$27_{20, }$ & 0 & 616  \\
C$_2$H$_5$CN & 251487.2 & $28_{17,}$-$27_{17,}$ & 1 & 780  \\
C$_2$H$_5$CN & 251501.0 & $28_{ 5,24}$-$27_{ 5,23}$ & 0 & 203  \\
C$_2$H$_5$CN & 251517.7 & $28_{ 8,}$-$27_{ 8,}$ & 1 & 539  \\
C$_2$H$_5$CN & 251520.6 & $28_{ 6,23}$-$27_{ 6,22}$ & 2 & 524  \\
C$_2$H$_5$CN & 251522.9 & $28_{ 6,22}$-$27_{ 6,21}$ & 2 & 524  \\
C$_2$H$_5$CN & 251558.1 & $28_{21, }$-$27_{21, }$ & 0 & 661  \\
C$_2$H$_5$CN & 251560.2 & $28_{ 6,23}$-$27_{ 6,22}$ & 1 & 509 \\
C$_2$H$_5$CN & 251561.2 & $28_{ 6,22}$-$27_{ 6,21}$ & 1 & 509  \\
C$_2$H$_5$CN & 251570.0 & $31_{ 6,26}$-$31_{ 5,27}$ & 0 & 253 \\
C$_2$H$_5$CN & 251607.1 & $28_{ 5,23}$-$27_{ 5,22}$ & 0 & 202  \\
C$_2$H$_5$CN & 251652.3 & $28_{ 5,24}$-$27_{ 5,23}$ & 2 & 511 \\
C$_2$H$_5$CN & 251661.0 & $10_{ 4, 7}$-$9_{ 3, 6}$ & 0 & 41  \\
C$_2$H$_5$CN & 251668.8 & $28_{ 4,25}$-$27_{ 4,24}$ & 0 & 193  \\
C$_2$H$_5$CN & 251691.0 & $28_{ 5,24}$-$27_{ 5,23}$ & 3 & 739  \\
C$_2$H$_5$CN & 251713.6 & $28_{ 5,23}$-$27_{ 5,22}$ & 2 & 511  \\
C$_2$H$_5$CN & 251728.7 & $10_{ 4, 6}$-$9_{ 3, 7}$ & 0 & 41  \\
\hline
\normalsize
\end{tabular}
\end{singlespace}
\newline
$^a$ Rotational energy levels are labeled with $J,K_a,K_c$, and omission of $K_c$ identifies a degenerate spectroscopic doublet in which $K_c=J-K_a$ and $K_c=J-K_a+1$. 
\newline
$^b$ Vibrational species: 0: ground state \citep{brauer09}, 1:~$ v_{13}=1$ \citep{kisiel20}, 2:~$v_{21}=1$ \citep{kisiel20}, 3:~$v_{20}=1$ \citep{daly13}.
\label{tab:lines}
\end{table}

In the next section the implications of the results for \etcyn\ are discussed.

\subsection{Discussion: \etcyn }

The ion \etcyn H$^+$ was inferred from early Cassini INMS mass spectra of Titan's upper atmosphere \citep[e.g.][]{vuitton07}, and the first identification of ethyl cyanide (propionitrile) in the neutral atmosphere was achieved using ALMA by \citet{cordiner15}. Nitrile molecules have a large molecular dipole ($\sim$4.0 D for small nitriles, equivalent to 70\% of an equivalent ionic bond), causing them to have strong rotational spectra. This was undoubtedly a reason why methyl cyanide (acetonitrile, \methcyn ) was first detected at sub-millimeter wavelengths \citep{bezard92}, and despite intensive searching has yet to be detected in the infrared \citep{nixon10b}. 

The detection of \etcyn\ by \citet{cordiner15} in ALMA Band 6 data at $\sim$ 220--240~GHz was close to the region observed in this work, with an abundance of $\sim$9 ppb (300 km step model) derived from disk-averaged observations in July 2012. Several years later, follow-up work by \citet{palmer17} also in Band 6 near 230~GHz determined a disk-average abundance of 7.2 $\pm$ 0.29 ppb for early 2014, while \citet{lai17} measured 7.37 $\pm$ 0.32 ppb from April 2015 data in Band 7 near 348~GHz. Our measurement of 8.5 $\pm$ 0.2 ppb (using a directly comparable 300 km step model) based on 2016 data falls in the mid-range of the previous measurements, and indicates that the global abundance was not changing substantially in this period. 


The vertical profile of propionitrile (as a global average) remains problematic. As pointed out by \citet{cordiner15} and also by \citet{lai17}, photochemical models typically over-estimate the abundance of this gas compared to retrieved abundances (see Fig.~\ref{fig:c2h5cn_profs} (b)), especially in the lower stratosphere, indicating a possible missing loss mechanism. The model which best replicates the data is that of \citet{willacy16} (Model C), which includes loss by condensation, sedimentation and haze formation. Despite the abundance at 100--300 km appearing significantly too high, the column abundance of ${\sim}10^{14}$ cm$^{-2}$ is of the same order of magnitude as our results. Other models \citep{loison15,vuitton19} substantially overestimate the column by a factor of ${\sim}100$, by continuing significant gas mixing fractions down to the tropopause.




\end{document}